\documentclass[letterpaper]{emulateapj}

\usepackage{epsfig}
\usepackage{amsmath}
\usepackage{rotating}
\usepackage{natbib}
\usepackage{lscape}
\usepackage{enumerate}
\usepackage{multirow}
\usepackage{array}

\def\plotone#1{\centering \leavevmode
\includegraphics[clip=, width=.85\columnwidth]{#1}}

\newcommand{\cN}[1]{\mathcal{N}}
\newcommand{\pn}[1]{\mbox{$(#1)$}}

\def\gsim{\;\rlap{\lower 2.5pt
 \hbox{$\sim$}}\raise 1.5pt\hbox{$>$}\;}
\def\lsim{\;\rlap{\lower 2.5pt
   \hbox{$\sim$}}\raise 1.5pt\hbox{$<$}\;}

\slugcomment{Submitted to ApJ}

\begin{document}


\title{
Models of Neptune-Mass Exoplanets: Emergent Fluxes and Albedos
}

\author{David S. Spiegel\altaffilmark{1},
Adam Burrows\altaffilmark{1},
Laurent Ibgui\altaffilmark{1},
Ivan Hubeny\altaffilmark{2},
John A. Milsom\altaffilmark{3}
}

\affil{$^1$Department of Astrophysical Sciences, Princeton University, Peyton Hall, Princeton, NJ 08544}

\affil{$^2$Steward Observatory, The University of Arizona, Tucson,
AZ 85721}

\affil{$^3$Department of Physics, The University of Arizona, Tucson,
AZ 85721}

\vspace{0.5\baselineskip}

\email{dsp@astro.princeton.edu, burrows@astro.princeton.edu,
ibgui@astro.princeton.edu, hubeny@as.arizona.edu,
milsom@physics.arizona.edu}

\begin{abstract}
There are now many known exoplanets with $M\sin i$ within a factor of
two of Neptune's, including the transiting planets GJ436b and
HAT-P-11b.  Planets in this mass-range are different from their more
massive cousins in several ways that are relevant to their radiative
properties and thermal structures.  By analogy with Neptune and
Uranus, they are likely to have metal abundances that are an order of
magnitude or more greater than those of larger, more massive planets.
This increases their opacity, decreases Rayleigh scattering, and
changes their equation of state.  Furthermore, their smaller radii
mean that fluxes from these planets are roughly an order of magnitude
lower than those of otherwise identical gas giant planets.  Here, we
compute a range of plausible radiative equilibrium models of GJ436b
and HAT-P-11b.  In addition, we explore the dependence of generic
Neptune-mass planets on a range of physical properties, including
their distance from their host stars, their metallicity, the spectral
type of their stars, the redistribution of heat in their atmospheres,
and the possible presence of additional optical opacity in their upper
atmospheres.
\end{abstract}


\section{Introduction}
\label{sec:intro}
Although in the early days of exoplanet research large, massive
planets (gas giants) were discovered in disproportionate numbers,
lower mass objects, including super Earths (1 to $\sim$10 times the
mass of the Earth) and ice-giants, are now being found with increasing
frequency.\footnote{See the catalog at http://exoplanet.eu.}  The
latter population, consisting of ``Neptune-mass'' bodies (those with
masses within a factor of a few of Neptune's), currently numbers
several dozen, and is growing rapidly.

Exoplanet transits, first observed by \citet{henry_et_al2000} and
\citet{charbonneau_et_al2000}, allow precise measurements both of
planet masses (by breaking the degeneracy between mass and inclination
angle) and of radii.  These two crucial pieces of data provide
constraints on models of structure, evolution, and bulk composition
\citep{ibgui+burrows2009, burrows_et_al2007, burrows_et_al2000,
guillot2008, guillot2005, guillot+showman2002, baraffe_et_al2005}.  As
a result, the best-studied among the Neptune-mass population are the
transiting planets GJ436b (1.33 times the mass of Neptune, and 1.27
times its radius; \citealt{caceres_et_al2009, bean+seifahrt2008,
bean_et_al2008, torres2007, deming_et_al2007b, maness_et_al2007}) and
HAT-P-11b ($1.50 M_{\rm Nep}$, $1.30R_{\rm Nep}$;
\citealt{bakos_et_al2009, dittman_et_al2009}).

Observations of an exoplanet can constrain theoretical models.  The
marriage of data with theory, therefore, allows us to learn about the
physical conditions on these distant worlds.  For example,
observations during transit can constrain various properties,
including both the atmospheric composition and structure at the
terminator \citep{seager+sasselov2000, hubbard_et_al2001,
charbonneau_et_al2002, desert_et_al2008, redfield_et_al2008} and the
motion of the atmosphere \citep{brown2001, spiegel_et_al2007b}.  This
paper, however, focuses on the diagnostics of planetary structure that
are available from observations of a planet's disk, not of its limb.
The disks of those planets whose orbits are aligned precisely enough
that they transit may be studied in exquisite detail.  Half an orbit
after transit, they undergo secondary eclipse, when they pass behind
their host stars.  Immediately before and after secondary eclipse, a
planet is in its ``full-moon'' phase, when its reflected and emitted
light are at their maximum from our vantage.  There have been a
variety of ground-based and space-based secondary-eclipse observations
of exoplanets in the last several years \citep{richardson_et_al2003,
snellen2005, deming_et_al2007b, deming_et_al2007c,
charbonneau_et_al2008, knutson_et_al2008b, knutson_et_al2009,
alonso_et_al2009a, alonso_et_al2009b, sing+morales-lopez2009}.  In the
last decade, several groups have produced a progression of theoretical
models of the atmospheres of Jupiter-mass exoplanets
\citep{sudarsky_et_al2000, sudarsky_et_al2003, hubeny_et_al2003,
burrows_et_al2005, fortney_et_al2006, burrows_et_al2008b,
fortney_et_al2008, burrows_et_al2008, spiegel_et_al2009b,
showman_et_al2009}, including calculations of both optical albedos and
infrared emergent spectra.  These calculations are now being used in
interpretative studies of the secondary eclipse
observations.\footnote{Comparable calculations have been made in the
context of Super-Earth planets \citep{miller-ricci_et_al2009a,
kaltenegger_et_al2009b}.}

A critical difference between the gas giant and the ice-giant planets
motivates the present investigation of theoretical models of lower
mass objects.  The gas giants of our solar system -- Jupiter and
Saturn -- have metal abundances not more than a few times solar
\citep{matter_et_al2009, saumon+guillot2004}.  However, Neptune,
Uranus, and, presumably, extrasolar planets in the same mass-range,
are thought to have bulk metallicities of 30-45 times solar
\citep{figueira_et_al2009, matter_et_al2009, guillot+gautier2007}.
Increased metallicity leads to increased opacity.  Reduced relative
hydrogen abundance might also reduce the importance of Rayleigh
scattering in the atmospheres of Neptune-mass planets.

In the last several years, infrared observations by the {\it Spitzer
Space Telescope}, in conjunction with theoretical models by our group
and others, have suggested that several giant planets have thermal
inversions, wherein, above a relative minimum, the atmospheric
temperature increases with altitude.  \citet{hubeny_et_al2003}
suggested that, if there is an additional source of upper atmosphere
optical opacity, incident stellar irradiation could lead to precisely
this type of atmospheric structure.  The planets whose emergent
spectra have suggested thermal inversions include HD 209458b
\citep{burrows_et_al2007c, knutson_et_al2008b}, HD 149026b
\citep{fortney_et_al2006, burrows_et_al2008b}, TrES-4
\citep{knutson_et_al2009}, XO-1b \citep{machalek_et_al2008}, XO-2b
\citep{machalek_et_al2009}, XO-3b \citep{mccullough_et_al2008}, and,
perhaps, $\upsilon$ Andromeda b \citep{burrows_et_al2008b}.  Various
authors have suggested that titanium oxide (TiO), as a strong optical
absorber, might provide the extra opacity that is needed to produce
the inferred inversions \citep{fortney_et_al2008, showman_et_al2009}.
However, \citet{spiegel_et_al2009b} argue that, without extremely
vigorous macroscopic mixing, a heavy molecule such as TiO will settle
to a level that is too deep for it to contribute to an inversion.
Furthermore, \citet{spiegel_et_al2009b} show that several planets for
which inversions have been inferred ought to have cold-trap regions
deeper in their atmospheres that are cool enough for titanium to
condense and rain out, therefore requiring even more vigorous mixing
if TiO is to survive up to the millibar levels where it would be
needed.\footnote{The analysis of \citealt{showman_et_al2009} appears
to agree with this conclusion.} \citet{zahnle_et_al2009} suggest that
sulfur photochemistry provides another avenue for achieving the
additional upper-atmosphere optical opacity that is needed to produce
inversions.  The ultimate cause of inversions remains unknown, and
might differ from one planet to another.  Nevertheless, sources of
extra optical opacity might be related to metal abundance, and so it
is reasonable to wonder whether some highly irradiated Neptune-mass
planets might have thermal inversions.

In order to survey the range of plausible structures of Neptune-mass
exoplanets, we compute a variety of model atmospheres.  We explore how
the emergent infrared fluxes and optical albedos of GJ436b and
HAT-P-11b depend both on the redistribution of heat in their
atmospheres and on the possible presence of an extra source of
atmospheric opacity (following the treatment in
\citealt{burrows_et_al2008b} and \citealt{spiegel_et_al2009b}).  We
also examine how ``ice-giant'' atmospheres, and their optical and
infrared fluxes, depend on their distances from their host stars,
their atmospheric metal abundances, and the spectral types of their
hosts.
By ``ice-giant,'' we are simply referring to a planet with
approximately Neptune's mass and radius; whether the atmosphere is
bounded below by ices is irrelevant to our analysis.  Our models are
of planets with hydrogen/helium-dominated atmospheres that are
metal-enriched relative to solar by a large factor.  We note that some
planets currently classified as ``Super-Earths'' might share this
structure.

The remainder of this paper is structured as follows: In
\S\ref{sec:methods}, we present our numerical techniques for computing
the atmospheres of ``ice-giant'' planets.  In \S\ref{sec:models}, we
describe the different models that we ran.  Section \ref{sec:results}
contains the results of our calculations.  Finally, in
\S\ref{sec:conc}, we summarize our findings.

\section{Numerical Methods}
\label{sec:methods}
We calculate radiative equilibrium models of irradiated planetary
atmospheres models.  As in our other recent studies, we use the code
{\tt COOLTLUSTY} \citep{hubeny_et_al2003, sudarsky_et_al2003,
burrows_et_al2006, burrows_et_al2008b, spiegel_et_al2009b}.  This code
is an variant of the code {\tt TLUSTY} \citep{hubeny1988,
hubeny+lanz1995}, with atomic and molecular opacities appropriate to
the cooler environments of planetary atmospheres and brown dwarfs
(\citealt{sharp+burrows2007, burrows+sharp1999, burrows_et_al2001},
augmented by \citealt{burrows_et_al2002} and
\citealt{burrows_et_al2005}).  Irradiation in planetary atmosphere
models is incorporated using Kurucz model stellar spectra,
interpolated to the temperatures and surface gravities appropriate to
our study \citep{kurucz1979,kurucz1994,kurucz2005}.

In order to examine the effect of metallicity on planet atmospheres,
we compute some models with solar abundance of metals, and others with
30$\times$solar abundance.  For both metallicities, we derive the
chemical equilibrium abundances (as functions of temperature and
density) of several hundred atomic and molecular species
\citep{sharp+burrows2007}.  Using chemical equilibrium abundances, we
calculate the monochromatic opacities that are used in the radiative
transfer calculations.  Furthermore, we recompute the equation of
state tables for the chemical abundances derived for 30$\times$solar
metallicity.

To treat the redistribution of incident stellar flux in a planet's
atmosphere, we use the $P_n$ formalism described in
\citet{burrows_et_al2006} and \citet{burrows_et_al2008b}.  In this
formalism, in lieu of calculating the full three dimensional general
circulation, the proportion of day-side incident flux that is
transported to and reradiated from the night-side is parameterized as
$P_n$, which plausibly ranges between 0, corresponding to all stellar
flux being instantaneously reradiated, and 0.5, corresponding to the
night side receiving approximately the same amount of stellar energy
as the dayside (as a result of advective heat redistribution).
The redistribution takes place between 0.003 bars and 0.6 bars.

For each model planet, we calculate the temperature-pressure profile,
the planet-star flux ratio of emergent infrared radiation, the optical
albedo, and the temperature and pressure, as functions of wavelength,
of the $\tau_\lambda=2/3$ photosphere.  It is worthwhile to note that,
since photons are not stamped as being reflected/scattered stellar
photons or emitted photons that arose in a planet's atmosphere, the
albedo is simply the quotient of emergent flux with incident flux
\citep{burrows_et_al2008,sudarsky_et_al2003,sudarsky_et_al2000}.
Defined this way, it is in principle possible for albedos to be
greater than 1.\footnote{In fact, this happens in the mid-infrared,
where planets are strong thermal emitters.}  However, at wavelengths
shorter than 1~$\mu$m, the albedos we report may be understood as
being predominantly measures of reflectance and scattering properties
of the day-side atmospheres.

\section{Planet Models}
\label{sec:models}
The known exoplanets in Neptune's mass range are found in a variety of
configurations.  Table~\ref{tab:neptunes_data} summarizes their
diversity.  Orbital separations range from 0.02~AU to 2.7~AU, and
stellar types from M through G are represented.  The diversity of
observed planet-star systems motivates our investigation of a
comparably diverse model set.

Our goal in this investigation is twofold: \pn{i} to examine the
diagnostics of the known transiting extrasolar ``ice-giants'' (GJ436b
and HAT-P-11b); and \pn{ii} to examine more generally the types of
radiative and thermal structures that might obtain in plausible
exo-Neptunian atmospheres.
%
%
To these ends, we compute 29 model atmospheres, listed in
Table~\ref{tab:models}.\footnote{There are 31 rows in
Table~\ref{tab:models}, of which two, the 26th and 30thre
duplicates.}$^,$\footnote{These models are publicly available at
http://www.astro.princeton.edu/$\sim$dsp/exoneptunes/ and at
http://www.astro.princeton.edu/$\sim$burrows/.}

Six of the models of GJ436b are named T35a03Z01P0k0G through
T35a03Z01P5k0G (corresponding to a GJ436b-like planet around a
GJ436-like star, with $P_n$ ranging between 0.0 and 0.5).\footnote{In
this labeling convention, the symbols have the following meanings:
``T'', for temperature, is followed by 0.01 times the star's effective
temperature; ``a'', for semimajor axis, is followed by 100 times the
orbital semimajor axis in AU; ``Z'', for metallicity, is followed by
the ratio with respect to solar abundance of metals; ``P'' is followed
by 10 times $P_n$; ``k'' is followed by 10 times $\kappa$ (in
cm$^2$~g$^{-1}$); the final letter (``G'' or ``H'') indicates whether
the radius and surface gravity of the model planet match those of
GJ436b or HAT-P-11b.}  Two additional models are T35a03Z01P3k2G
(corresponding to GJ436b with $P_n=0.3$ and an additional optical
absorber of opacity $\kappa = 0.2\rm~cm^2~g^{-1}$ in the upper
atmosphere, where $\kappa$ is the same as $\kappa_e$ from
\citealt{burrows_et_al2008b}) and T35a03Z30P0k0G (corresponding to
GJ436b with $P_n=0.0$ and no extra optical absorber, and with opacity
corresponding to 30$\times$solar metallicity).  As in previous work,
$\kappa$ is gray over the range $3\times 10^{14}$-$7\times
10^{14}$~Hz.  The models of HAT-P-11b are T48a05Z01P0k0H through
T48a05Z01P5k0H (HAT-P-11b with $P_n$ ranging from 0.0 to 0.5) and
T48a05Z01P3k2H (extra upper atmosphere absorber of opacity $\kappa =
0.2\rm~cm^2~g^{-1}$).  The additional optical opacity in models
T35a03Z01P3k2G and T48a05Z01P3k2H is confined to the pressure range
from $\sim$5~mbar to the top of the atmosphere.

We also test more generic ``ice-giant'' models, broadly sampling a
wide range of possible conditions.  In particular, we test the effect
of orbital separation, of metallicity, and of stellar type.
OGLE-05-169L-b, at 2.7~AU, is a bit of an outlier and will not be
observed in the near future. The remainder of the known Neptune-mass
exoplanets are predominantly at separations $a \lsim 0.5$~AU.  This
motivates our decision to examine planets in this range of separations
in this initial study.  Models testing the effects of orbital
separation and of metallicity are T58a03Z01P1k0G through
T58a50Z01P1k0G (solar metallicity models from 0.03~AU to 0.50~AU
around a sun-like star), and T58a03Z30P1k0G through T58a50Z30P1k0G
(30$\times$solar metallicity, at the same distances from the same
star).  Models testing the effect of stellar type are T35a03Z01P1k0G
through T62a03Z01P1k0G.  These models place a GJ436b-size planet at
0.03~AU from, respectively, GJ436 ($T=3500$~K, $\log_{10}g = 4.80$),
an M0V star ($T=4200$~K, $\log_{10}g = 4.70$), HAT-P-11 ($T=4780$~K,
$\log_{10}g = 4.60$), a KOV star ($T=5200$~K, $\log_{10}g = 4.50$), a
sun-like star ($T=5778$~K, $\log_{10}g = 4.44$), and an F9V star
($T=6200$~K, $\log_{10}g = 4.25$).

\section{Results}
\label{sec:results}

For each model planet, we present two observable variables and three
that are not observable.  The observables are the emergent infrared
flux, presented as a planet-star flux ratio, and the optical albedo.
The three variables that are unobservable, but are nonetheless useful
probes of our models, are the following: the temperature-pressure
profile (temperature as a function of depth in the atmosphere); the
temperature, as a function of wavelength, of the $\tau_\lambda=2/3$
photosphere; and the pressure, as a function of wavelength, of the
same photosphere.

\subsection{GJ436b and HAT-P-11b}
\label{ssec:gj436b}

\begin{figure}
\plotone
{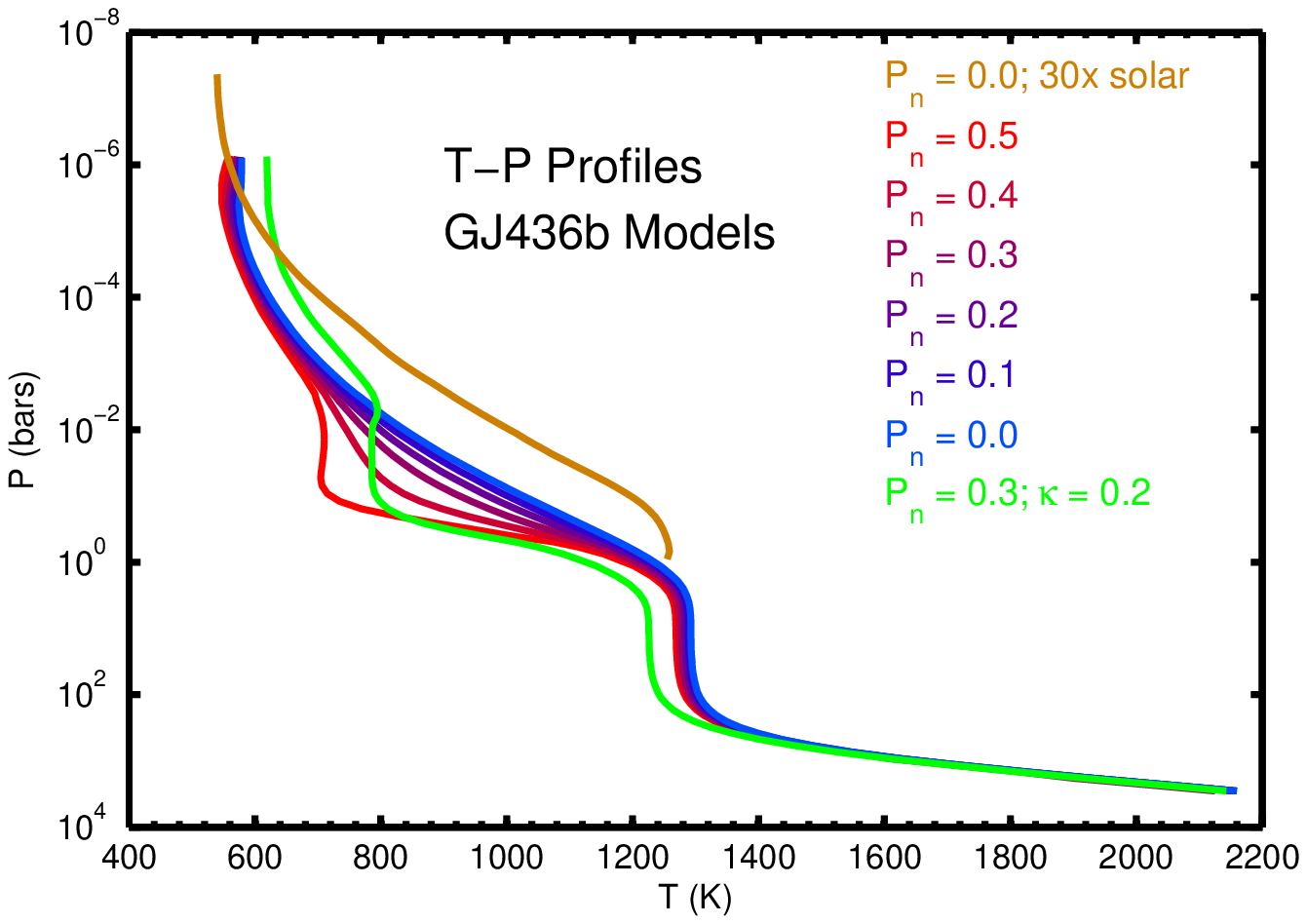}
\plotone
{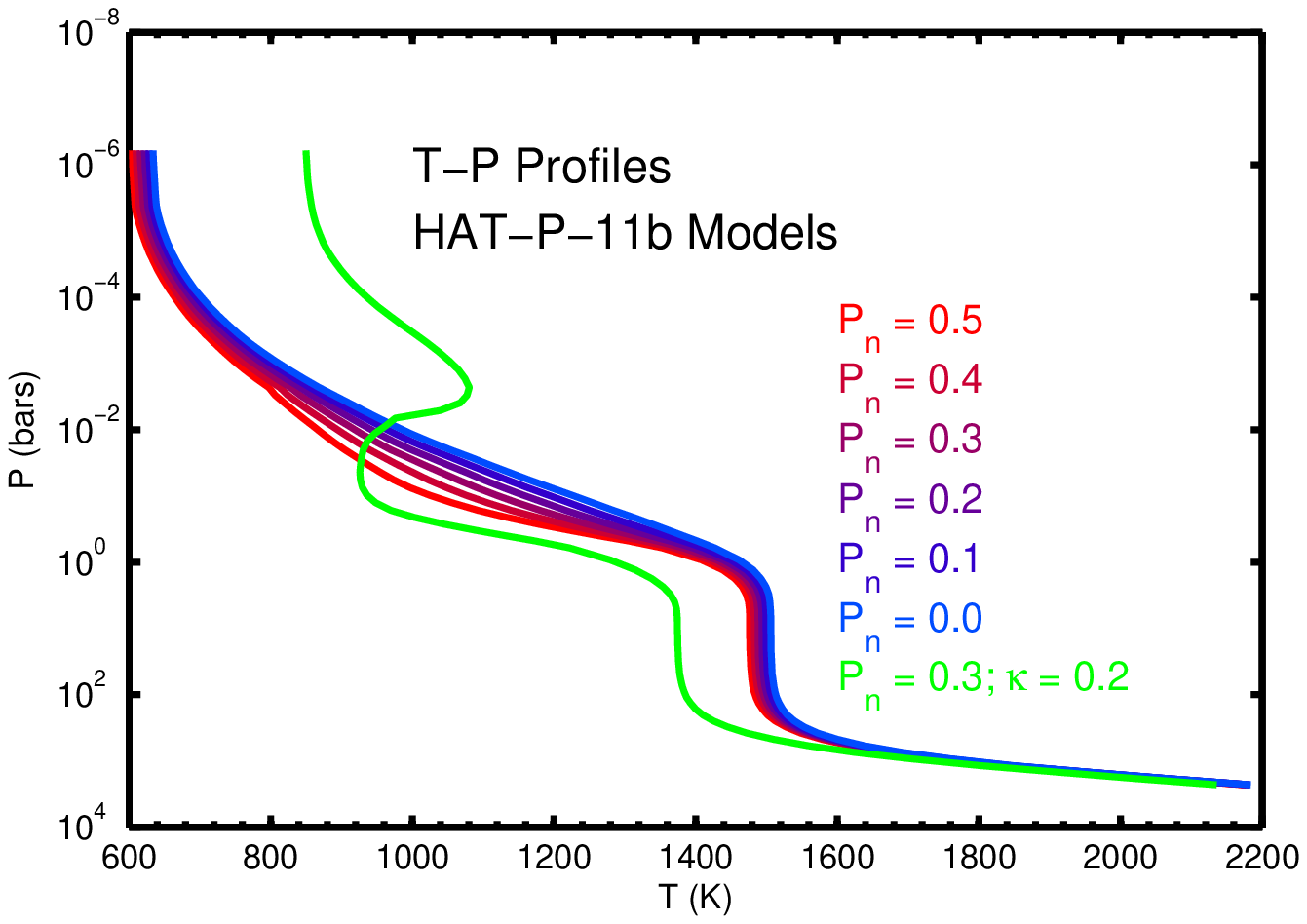}
\caption{Model temperature-pressure profiles of GJ436b and HAT-P-11
b.  {\it Top:} Models of GJ436b, with the redistribution parameter
$P_n$ varying from 0 (no redistribution) to 0.5 (complete
redistribution of heat to the night side).
A model is shown with added upper atmosphere opacity $\kappa =
0.2\rm~cm^2~g^{-1}$ (here, $\kappa$ is the same as the $\kappa_e$ from
\citealt{burrows_et_al2008b}).  An additional model shows the
  influence of 30$\times$solar atmospheric opacity. {\it Bottom:}
Same, for HAT-P-11 b, except without the 30$\times$solar model.}
\label{fig:Pn_prof}
\end{figure}

The most salient difference between GJ436b and HAT-P-11b is that the
latter orbits a hotter, brighter star.  The temperature-pressure
profiles of models of them, portrayed in Fig.~\ref{fig:Pn_prof},
reflect this: HAT-P-11b (bottom panel of Fig.~\ref{fig:Pn_prof}) is
consistently hotter at a given pressure.  For both profiles, the
redistribution of day-side irradiation to the night side reduces the
dayside's temperature.  As $P_n$ increases, the thermal profile
becomes cooler at a given pressure in the range where the
redistribution takes place (0.003~bars to 0.6~bars in these models).
This effect is more pronounced in GJ436b than in HAT-P-11b.

On the other hand, the effect of an additional optical absorber (green
curves, $\kappa = 0.2\rm~cm^2~g^{-1}$) is more pronounced in
HAT-P-11b's profile -- precisely because its star is hotter and,
therefore, brighter in the optical.  In GJ436b's model, an additional
optical absorber at high altitude increases upper atmosphere
temperatures by a modest $\sim$60~K.  In HAT-P-11b's model, however,
an extra optical absorber increases the temperature at
$\sim$10$^{-3}$~bars by $\sim$300~K, and creates a thermal inversion
of about 100~K.

In the figure for GJ436b, we have included one additional model
(T35a03Z01P0k0G), with $P_n=0.0$ and 30$\times$solar opacity.  The
profile of this model looks qualitatively different from those of the
other models, because it does not quite extend to the
marginally-stable convection zone.  However, the Rosseland mean
optical depth at the base of this model is $\sim$80, and the net flux
is constant with altitude, so the unmodeled portions of the atmosphere
do not affect the emergent spectrum.

\begin{figure}[ht]
\centerline{
\includegraphics[width=10cm,height=8.4cm,angle=0,clip=true]{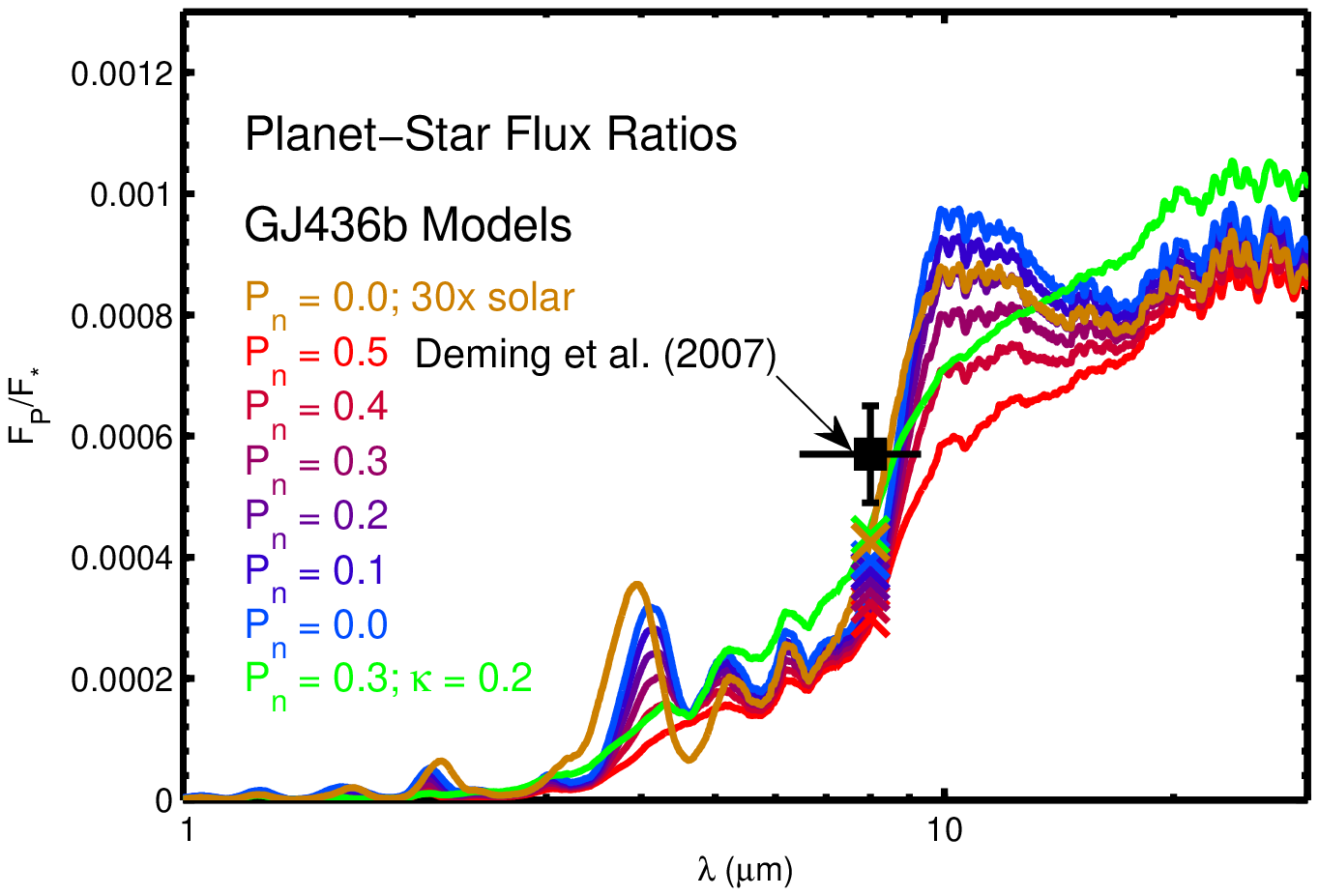}
\includegraphics[width=10cm,height=8cm,angle=0,clip=true]{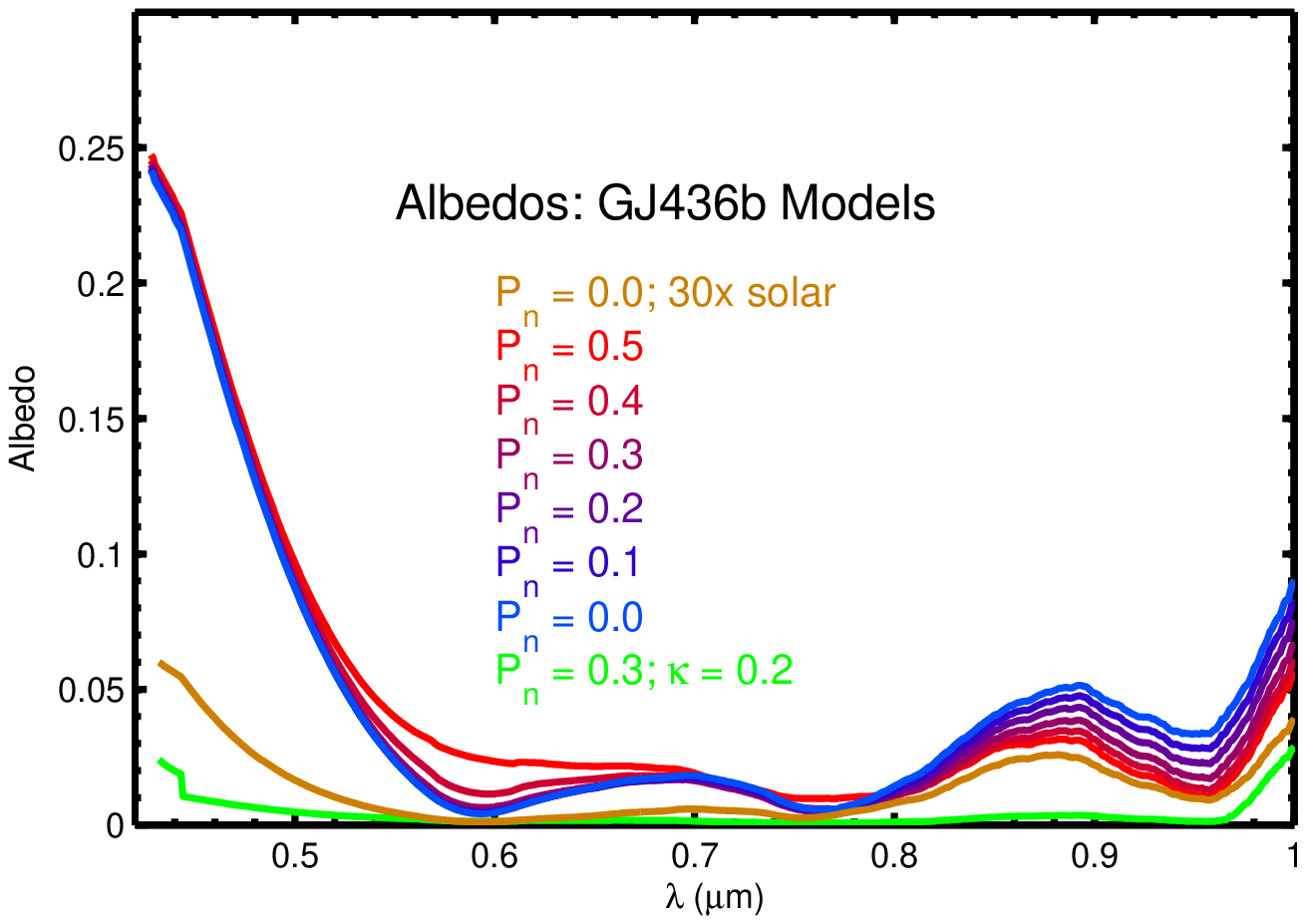}
}
\centerline{
\includegraphics[width=9.5cm,height=8cm,angle=0,clip=true]{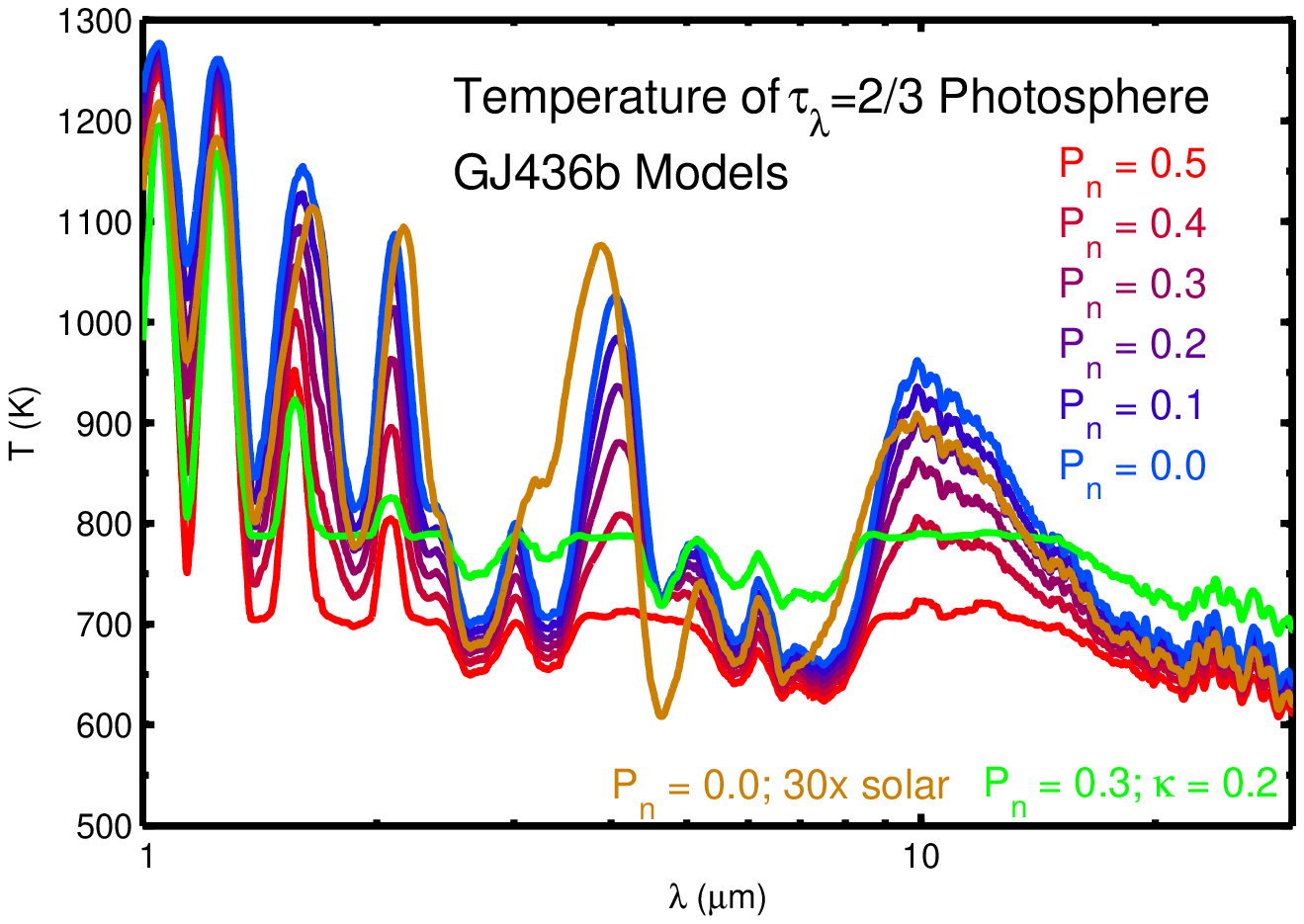}
\includegraphics[width=9.8cm,height=8cm,angle=0,clip=true]{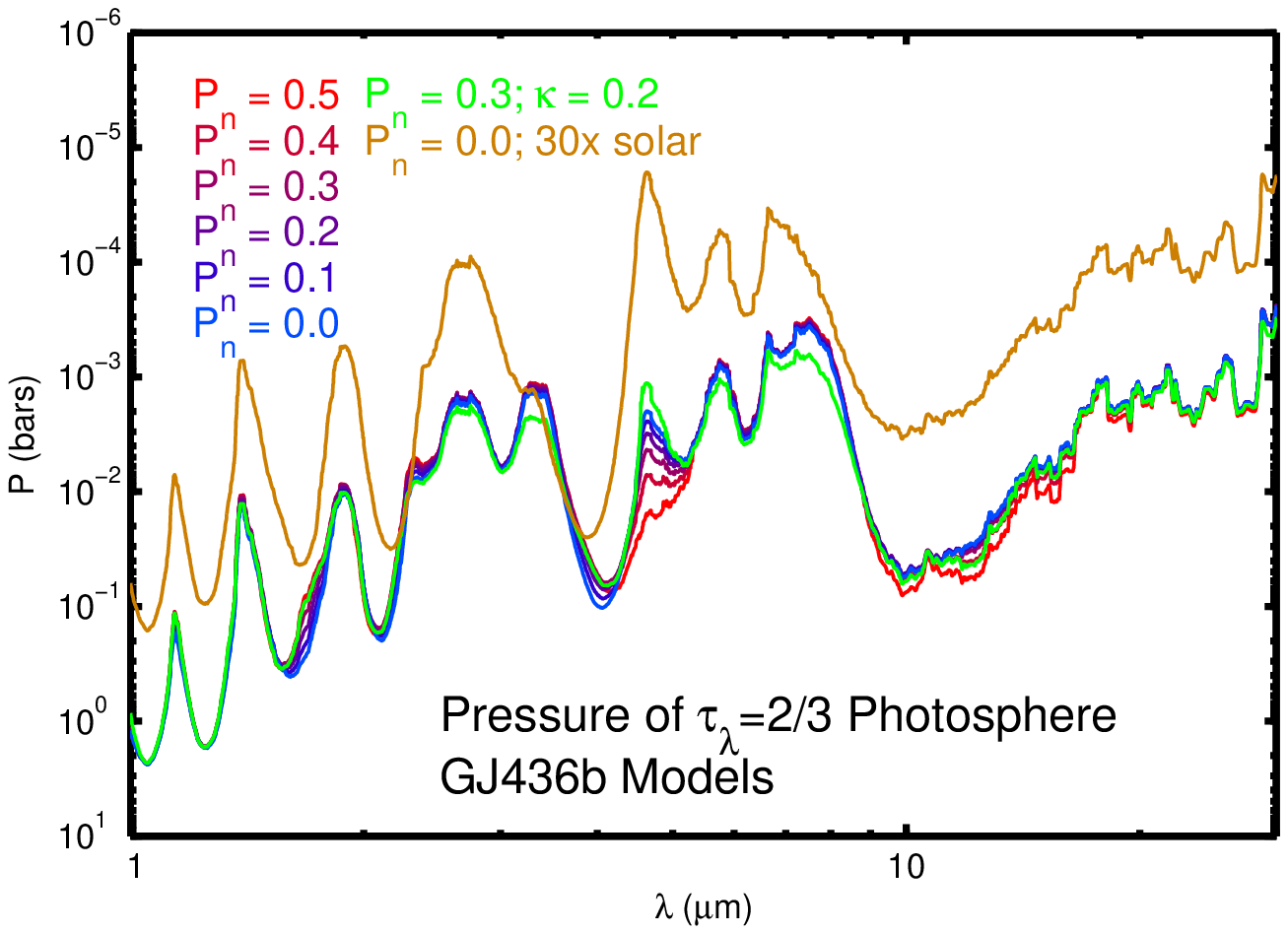}
}
\caption{Additional characteristics of models of GJ436b depicted in
the top panel of Fig.~\ref{fig:Pn_prof}.  {\it Top left:} Planet-star
flux ratios. Varying $P_n$ has little influence on the planet's flux
at 8~$\mu$m, but an additional upper atmosphere absorber results in a
significantly better match of the model to the data.
30$\times$solar atmospheric opacity also helps bridge the gap
between the solar-opacity models and the observed data.
The black square is the {\it Spitzer} IRAC-4 measurement of
\citet{deming_et_al2007b}; vertical bars indicate 1-$\sigma$
uncertainty; the horizontal bar indicates the full width at 10\%
maximum of the IRAC-4 band's transmission function.  Large X's
indicate integrated planet flux divided by integrated stellar flux
over this band.
{\it Top right:} Albedos. The albedo is lower for lower $P_n$, and
much lower for nonzero $\kappa$.  {\it Bottom left:} Temperature of
the $\tau_\lambda=2/3$ surface, as a function of wavelength. {\it
Bottom right:} Pressure of the $\tau_\lambda=2/3$ surface, as a
function of wavelength.}
\label{fig:GJ436Pn_spec_alb_etc}
\end{figure}

\begin{figure}
\centerline{
\includegraphics[width=10cm,height=8.4cm,angle=0,clip=true]{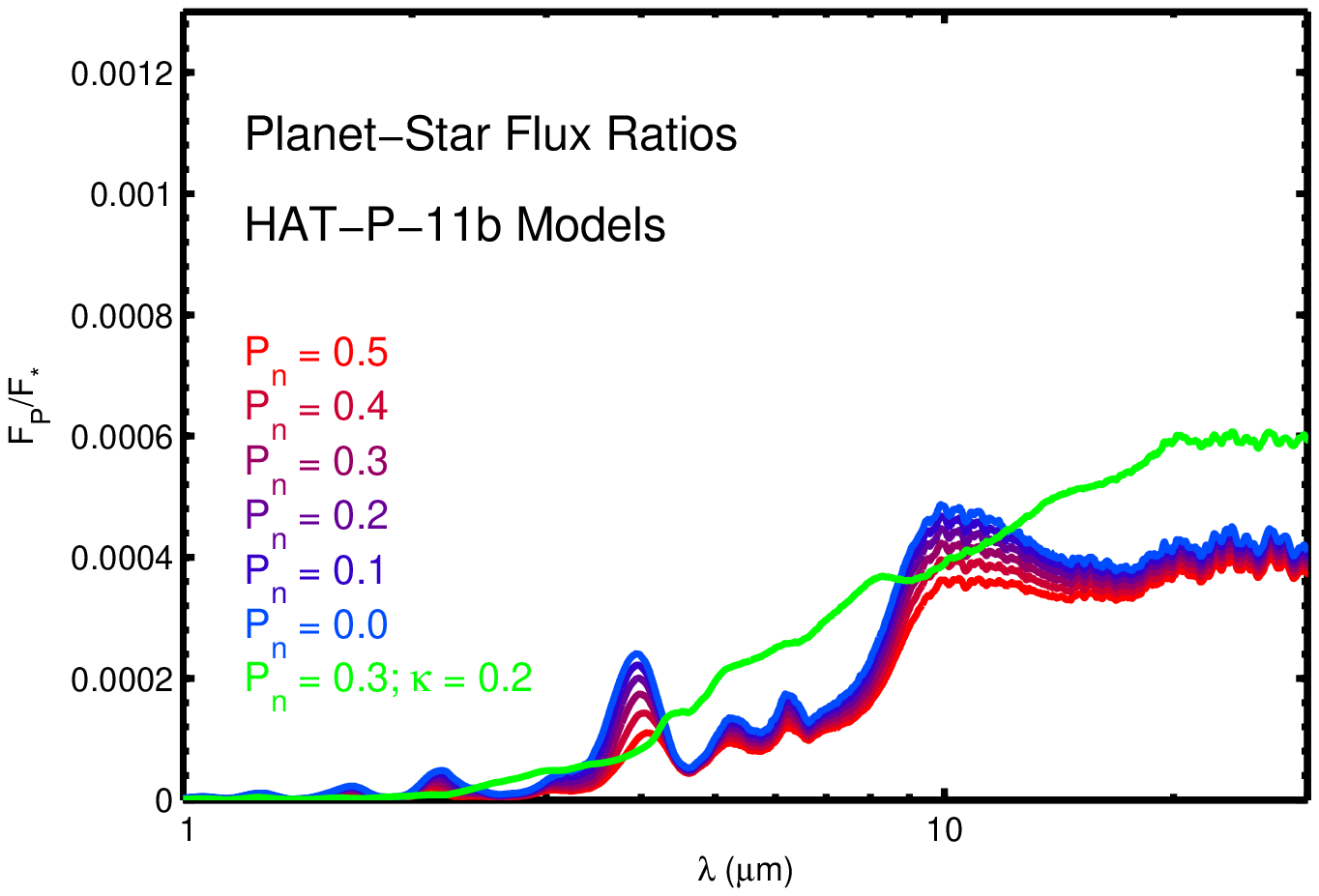}
\includegraphics[width=10cm,height=8cm,angle=0,clip=true]{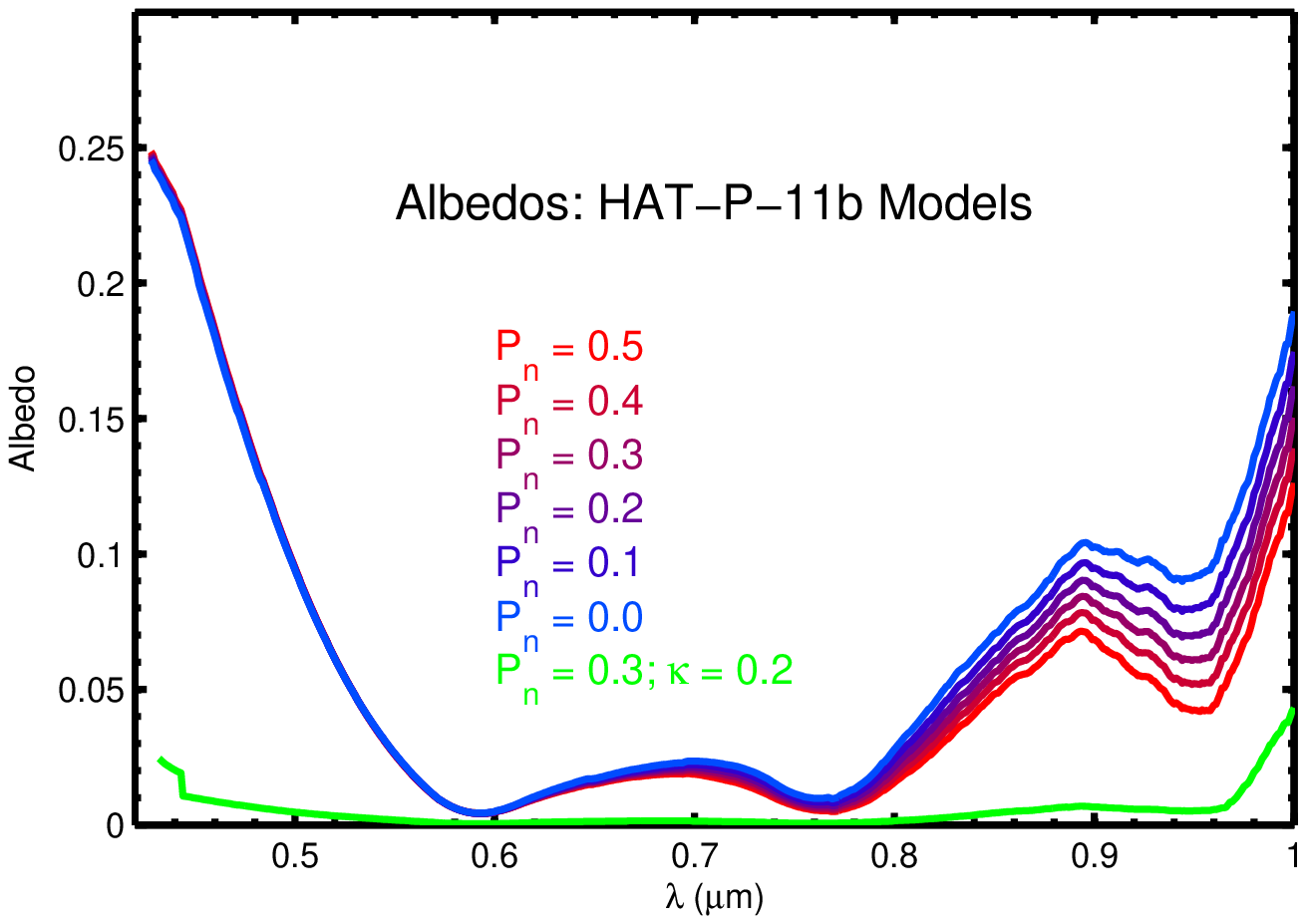}
}
\centerline{
\includegraphics[width=9.5cm,height=8cm,angle=0,clip=true]{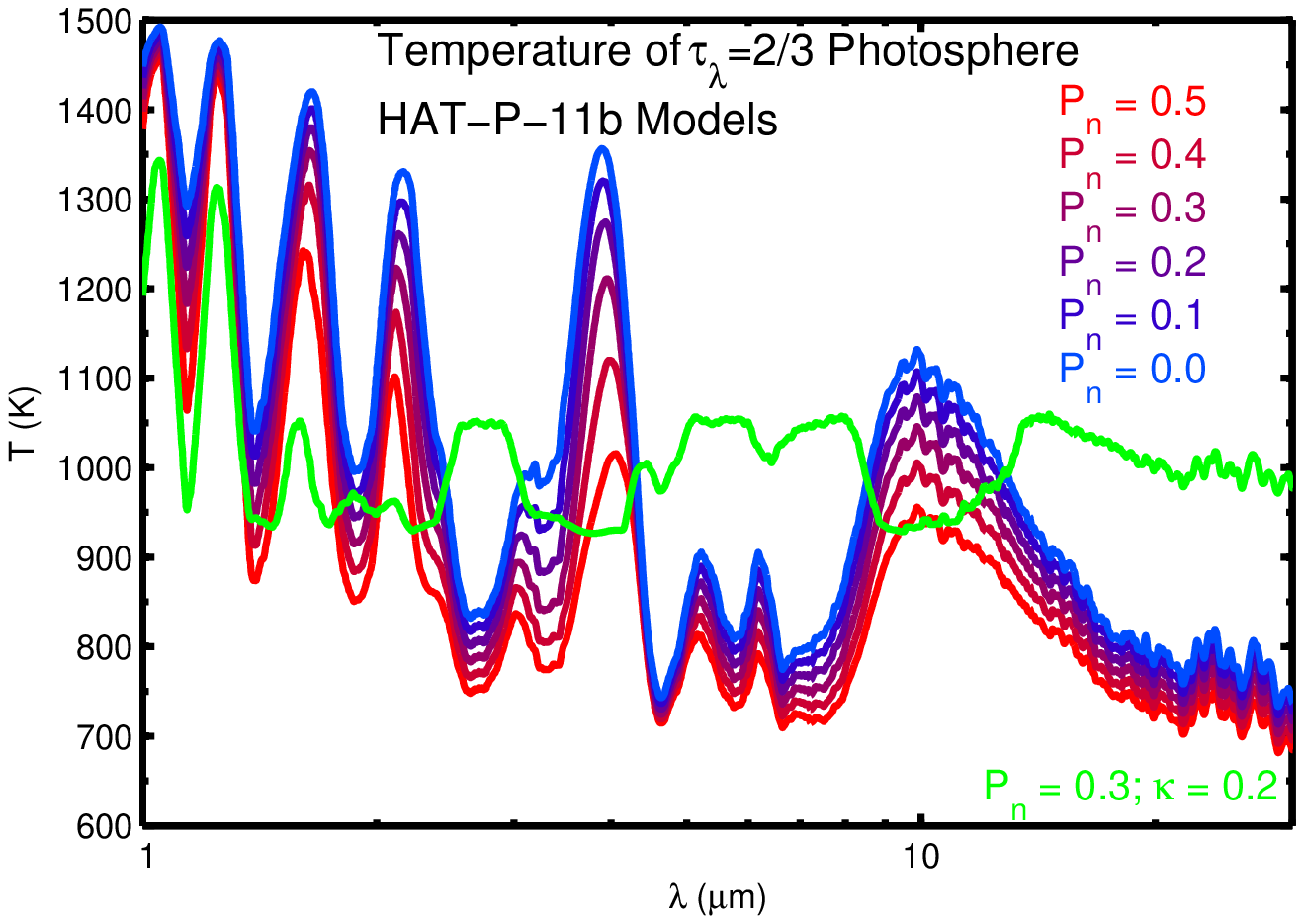}
\includegraphics[width=9.8cm,height=8cm,angle=0,clip=true]{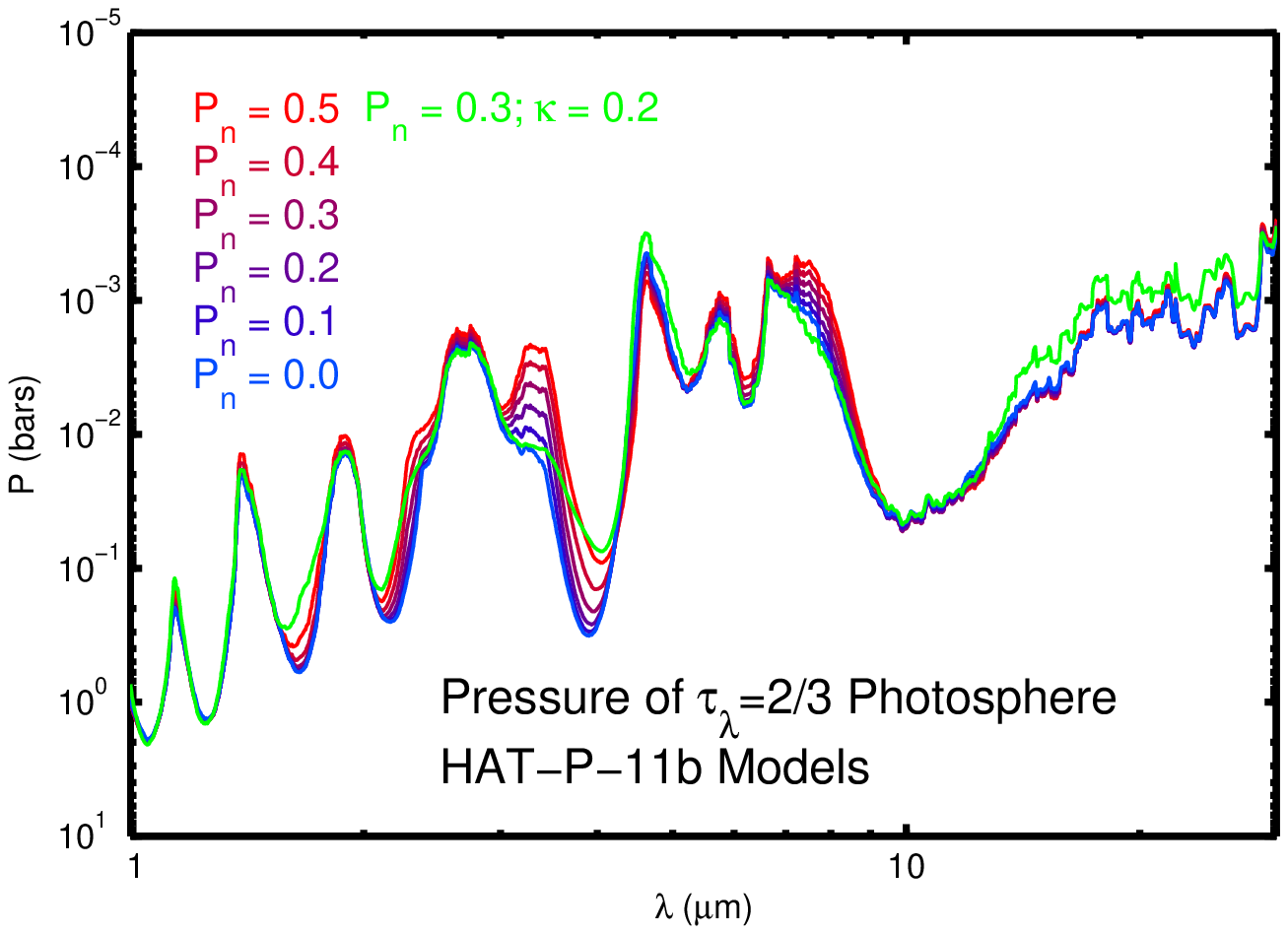}
}
\caption{Additional characteristics of models of HAT-P-11b depicted in
the bottom panel of Fig.~\ref{fig:Pn_prof}.  Completely analogous to
Fig.~\ref{fig:GJ436Pn_spec_alb_etc}.}
\label{fig:HP11Pn_spec_alb_etc}
\end{figure}

Let us consider our four wavelength-dependent model diagnostics,
presented in Fig.~\ref{fig:GJ436Pn_spec_alb_etc} (GJ436b) and
Fig.~\ref{fig:HP11Pn_spec_alb_etc} (HAT-P-11b).  We have smoothed the
curves in each plot to a spectral resolution of 100 for ease in
viewing.  In the flux-ratio plot of
Fig.~\ref{fig:GJ436Pn_spec_alb_etc} (upper left panel), we include the
8-$\mu$m measurement that \citet{deming_et_al2007b} obtained with the
{\it Spitzer} InfraRed Array Camera (IRAC).

The models' emergent infrared fluxes show various spectral features.
The opacity database contains significant CH$_4$ cross section at
3.3~$\mu$m and 7.8~$\mu$m, H$_2$O and CO absorption at 4.5~$\mu$m, and
H$_2$O at 6~$\mu$m and 10~$\mu$m \citep{sharp+burrows2007}.  The
emergent flux is insensitive to $P_n$ at some wavelengths, and shows a
moderate-to-strong dependence on $P_n$ at others (stronger in the
models of GJ436b than in those of HAT-P-11b).  The effect of
increasing $P_n$ is to reduce the emergent flux in ``emission''
features, specifically at $\sim$4~$\mu$m, $\sim$5~$\mu$m,
$\sim$6.5~$\mu$m, and $\sim$10-15~$\mu$m.

The optical albedo (upper right panels of both figures) does not
exhibit a strong dependence on $P_n$.  At wavelengths shorter than
0.5~$\mu$m, the albedo is nonnegligible, due to Rayleigh scattering.
At longer wavelengths, it tends to be $\lsim$0.05.  In both figures,
the albedos of models with $\kappa=0.0\rm~cm^2~g^{-1}$ start to
increase longward of $\sim$0.8~$\mu$m.  Albedos of HAT-P-11b models
are somewhat higher at these wavelengths.  Note, for both models, the
dip in optical albedo at $\sim$0.95~$\mu$m, which is due to a water
absorption feature \citep{sharp+burrows2007}.

Although the temperature and pressure of the $\tau_\lambda=2/3$
photosphere are not observable (bottom left and right panels,
respectively), it is instructive to examine the effect of increasing
$P_n$ on these variables.  When $P_n=0.0$, the temperature varies
significantly with wavelength, as narrow windows of low opacity allow
flux from the deeper, warmer portions of the atmosphere to escape.  As
$P_n$ increases, the location of the photosphere becomes more nearly
constant with wavelength.  The pressure of the photosphere does not
vary much with $P_n$, except near 5~$\mu$m.  As seen in the bottom
right panel, most of the infrared spectrum originates at pressures
between 10$^{-3}$ and 0.1 bars.  For these pressures, as seen in the
upper left panel, the day-side energy sink for $P_n=0.5$ (and to a
lesser extent for lower $P_n$'s) decreases the $T$ gradient so that
for $P_n=0.5$ there is a quasi-plateau in temperature plateau these
pressures.

The effect of an added source of opacity in the high atmosphere has
striking consequences for both the emergent flux and albedo.  Although
models T35a03Z01P0k0G through T35a03Z01P5k0G
($\kappa=0.0\rm~cm^2~g^{-1}$) all produce lower planet flux at
8~$\mu$m than \citet{deming_et_al2007b} measured, model T35a03Z01P3k2G
($\kappa=0.2\rm~cm^2~g^{-1}$) has an IRAC-bandpass-integrated 8-$\mu$m
flux that is more than 20\% higher than that of T35a03Z01P3k0G (the
$P_n=0.3$ model with no extra optical absorber), and is closer to
being consistent with the data point's error bars.\footnote{A model
with $\kappa=0.5\rm~cm^2~g^{-1}$ (not shown) passes right through the
center of the error bars.}  Furthermore, model T35a03Z30P0k0G, with
zero redistribution and 30$\times$solar opacity, has an integrated
8-$\mu$m flux that is 7.5\% higher than that of the analogous model
with solar opacity.  Still, its 8-$\mu$m flux is 12\% lower than
bottom of the 1-$\sigma$ range on the \citet{deming_et_al2007b}
measurement.  In both models T35a03Z01P3k2G and T48a05Z01P3k2H (GJ436b
and HAT-P-11b with $\kappa=0.2\rm~cm^2~g^{-1}$), an extra optical
absorber causes the upper atmosphere to be warmer and the planet flux
and the photospheric temperature to show fewer spectra features.
Despite the warmer upper atmosphere temperatures and generally warmer
photospheres, however, the additional optical opacity causes the the
albedo to be much lower in the optical ($\lsim$0.01 throughout much of
the optical) because Rayleigh scattering is diminished.  This is the
case even though the thermal component of emission is greater by dint
of the higher temperature.\footnote{At sufficiently high values of
$\kappa$ and with sufficient irradiation, this trend would reverse,
and higher $\kappa$ would lead to a larger apparent albedo.}

Consider again the flux-ratio comparisons for GJ436b (upper left
panels of Fig.~\ref{fig:GJ436Pn_spec_alb_etc}).  All 8 models have
approximately the same flux at 4.6~$\mu$m, but flux varies
significantly at 4.0~$\mu$m.  The ratio of the planet's flux at
4.0~$\mu$m to that at 4.6~$\mu$m varies from $\sim$2.4 for
T35a03Z01P0k0G to $\sim$1.2 for T35a03Z01P4k0G and T35a03Z01P3k2G to
$\sim$0.8 for T35a03Z01P5k0G.  Similarly, the ratio of flux at
$\sim$10~$\mu$m to flux at 4.6~$\mu$m provides a constraint on the
redistribution of heat.  This ratio varies from $\sim$7.6 for
T35a03Z01P0k0G to $\sim$4.5 for T35a03Z01P5k0G.  These flux ratios,
therefore, could provide a handle on the amount of redistribution of
heat.  In addition, at 7 to 8~$\mu$m, extra upper atmosphere opacity
(T35a03Z01P3k2G) increases the emergent flux to $\sim$20\% above what
it would be without the extra absorber.  We note that the difference
between the 30$\times$solar opacity model and those with solar opacity
appears greater when one considers the full spectral shape than when
integrated over the 8-$\mu$m IRAC band.

The models of HAT-P-11b exhibit a qualitatively analogous dependence
on $P_n$, although the ratios are quantitatively somewhat different
because of the differences between the two stars and the two orbital
separations.  But, again, T48a05Z01P0k0H has significantly greater
flux at 4.0~$\mu$m (by a factor of $\sim$2) and at 10~$\mu$m (by a
factor of $\sim$1.4) than T48a05Z01P5k0H.  Furthermore, because
HAT-P-11b is a hotter star with correspondingly more of its flux in
the optical, an extra optical absorber in the upper atmosphere has a
more significant effect on this planet than on GJ436b.  Between
7-8~$\mu$m, T48a05Z01P3k2H is approximately twice as bright as the
models without an extra absorber, and is also much brighter (by
$\sim$40\%) at wavelengths greater than 15~$\mu$m.

\subsection{Distance}
\label{ssec:dist}
The temperature of an irradiated stellar companion's photosphere drops
off rapidly with increasing orbital separation, and its emergent flux
decreases correspondingly.  For separations greater than $\sim$0.5~AU,
both the reflected/scattered optical light and the infrared emission
become too dim to be detected without spatially separating (and
masking) the light of the star.  Although direct detection of
wide-separation Neptune-mass objects will be feasible in the future
\citep{burrows2005}, in this paper we focus on objects within 0.5~AU
of their primaries.

The top panel of Fig.~\ref{fig:dist_Z_stars_prof} presents
temperature-pressure profiles of models at 0.03, 0.05, 0.10, 0.30, and
0.50~AU from a sun-like star (G2V).  In these models (T58a03Z01P0k0G
through T58a50Z30P0k0G in Table~\ref{tab:models}),
$\kappa=0.0\rm~cm^2~g^{-1}$ and $P_n=0.0$.  Solar opacity models are
represented by solid curves, and 30$\times$solar opacity models with
dashed curves; the differences between these are discussed in
\S\ref{ssec:metal} below.

Figure~\ref{fig:dist_Z_spec_alb} contains four panels analogous to the
four of Fig.~\ref{fig:HP11Pn_spec_alb_etc}.  The predominant trends
with distance in the flux-ratio plot (upper left panel) and in the
photospheric temperature plot (lower left panel) are, respectively,
that both the planet flux and planet temperature decrease as the model
planet is moved farther from the star.  In addition, there is some
change in the spectral shape -- cooler, more distant planets have
redder spectra.  The albedos of these models (upper right panel of
Fig.~\ref{fig:dist_Z_spec_alb}) change from low at short wavelengths
and high ($\sim$0.7) at long wavelengths for the 0.03~AU models to
moderate ($\sim$0.3) at short wavelengths and low at long wavelengths
for the 0.50~AU models.  The close-in models have relatively high
apparent albedo at wavelengths approaching 1~$\mu$m because of the
contribution from the Wien tails of their thermal emission.

\subsection{Metallicity}
\label{ssec:metal}
\begin{figure}
\plotone
{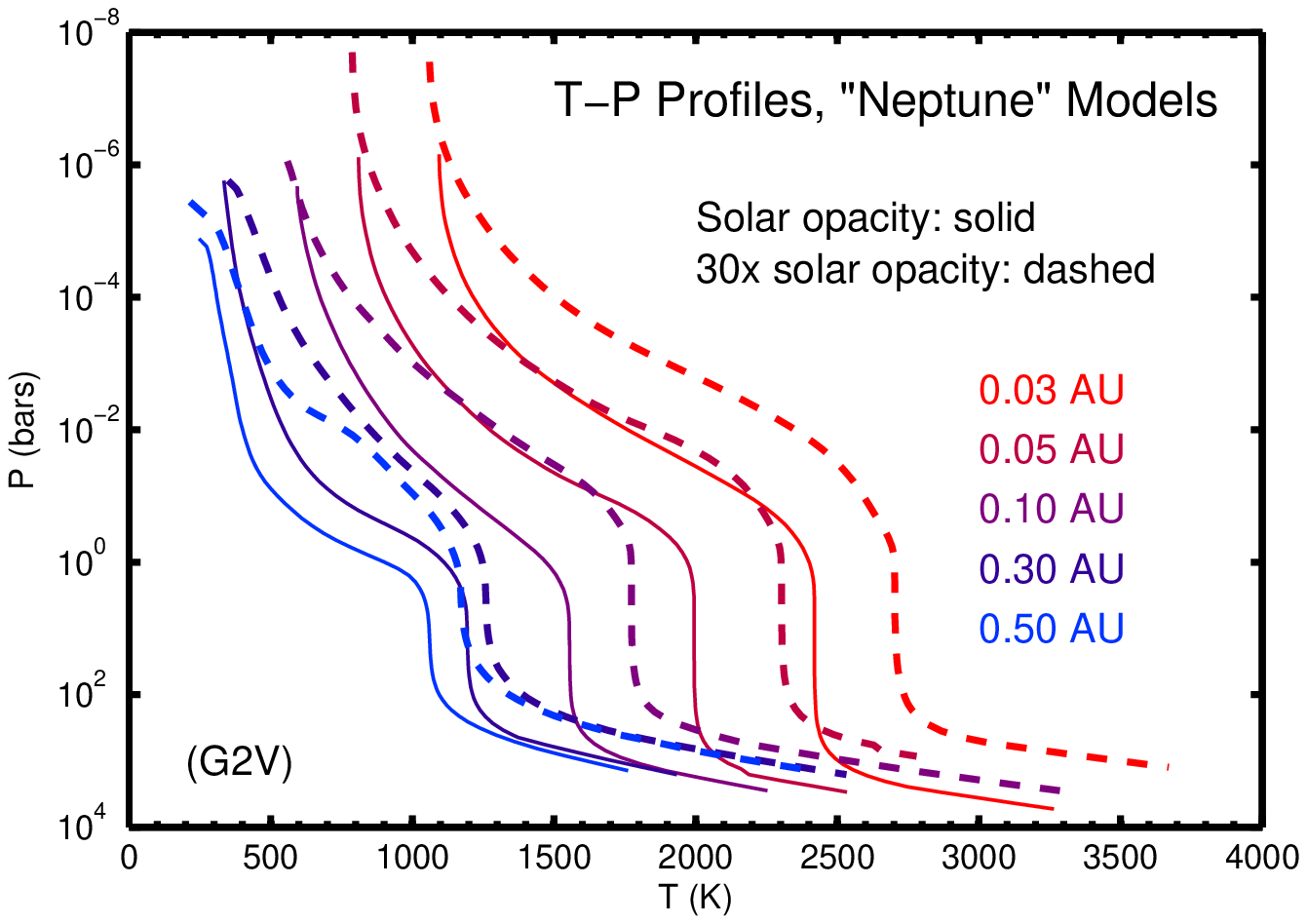}
\plotone
{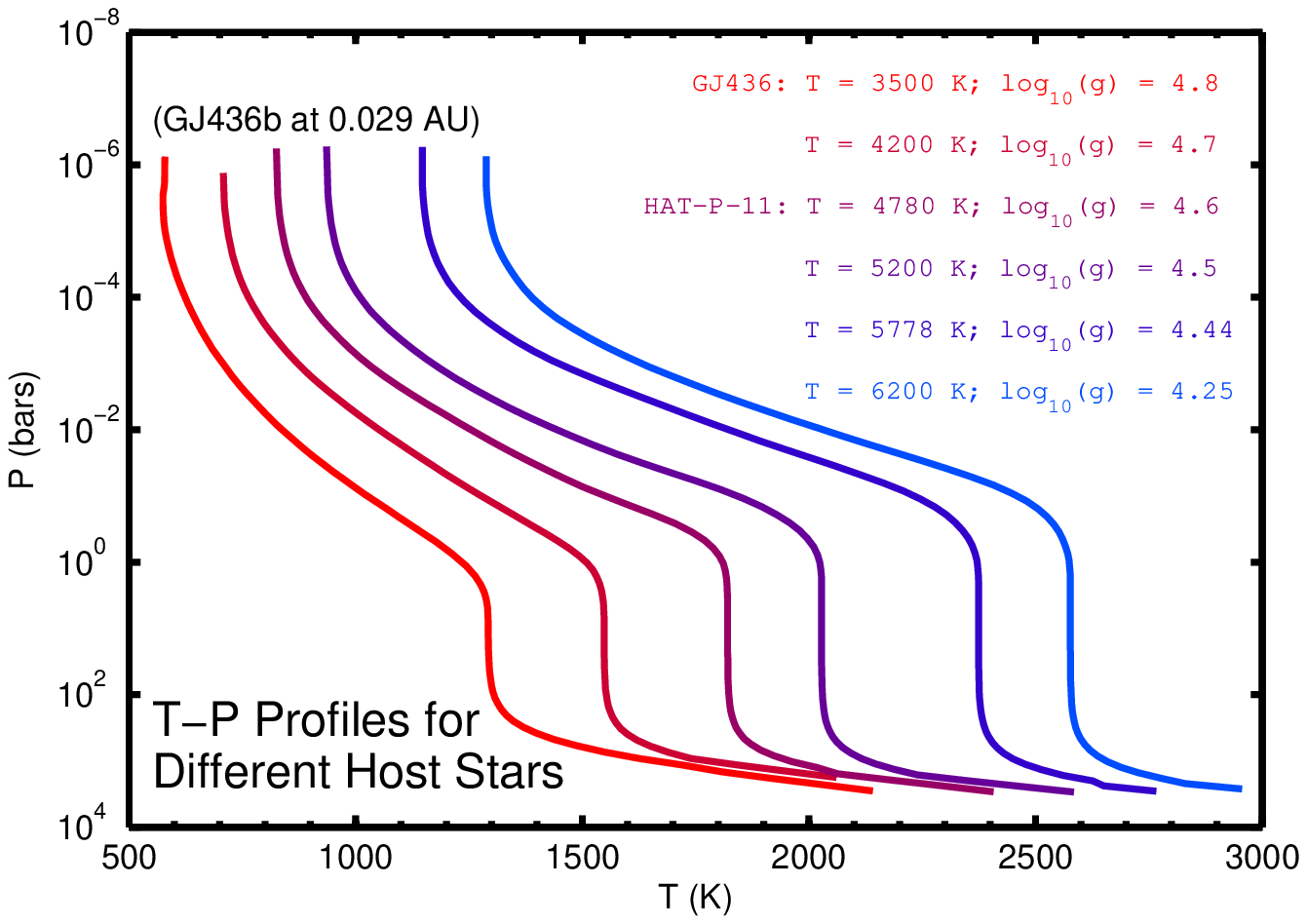}
\caption{Model temperature-pressure profiles of GJ436b-like planet.
In all models, the redistribution parameter $P_n = 0$. {\it Top:}
Models are computed at a range of distances from a Sun-like star, at
solar and at 30-times solar metallicity.  More distant models are
cooler, and more metal-rich models are warmer at a given
pressure. {\it Bottom:} Models are computed for a range of host stars,
from GJ436 (a cool M-dwarf) through an F9 star (6200~K).}
\label{fig:dist_Z_stars_prof}
\end{figure}

\begin{figure}
\centerline{
\includegraphics[width=10cm,height=8.4cm,angle=0,clip=true]{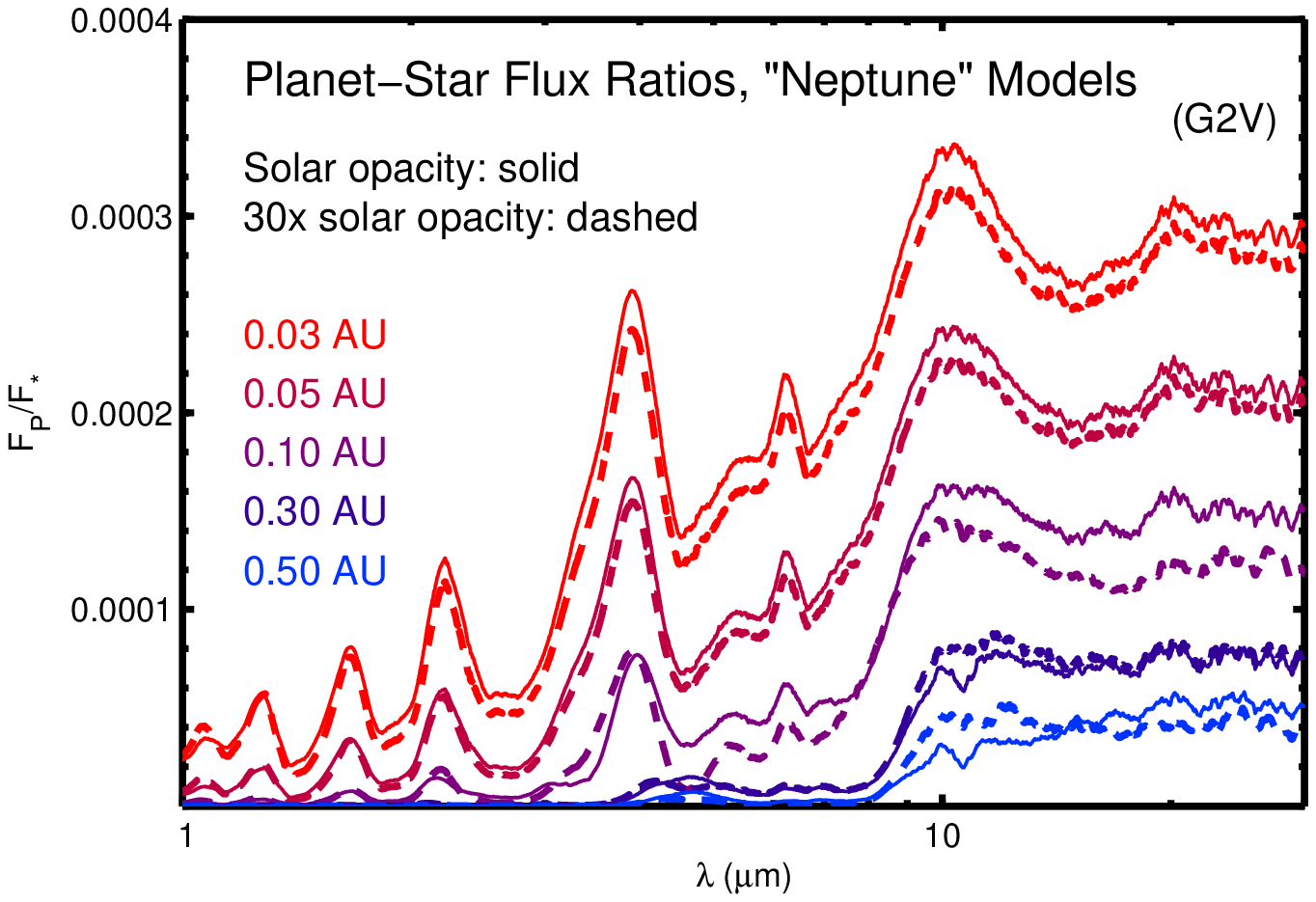}
\includegraphics[width=10cm,height=8cm,angle=0,clip=true]{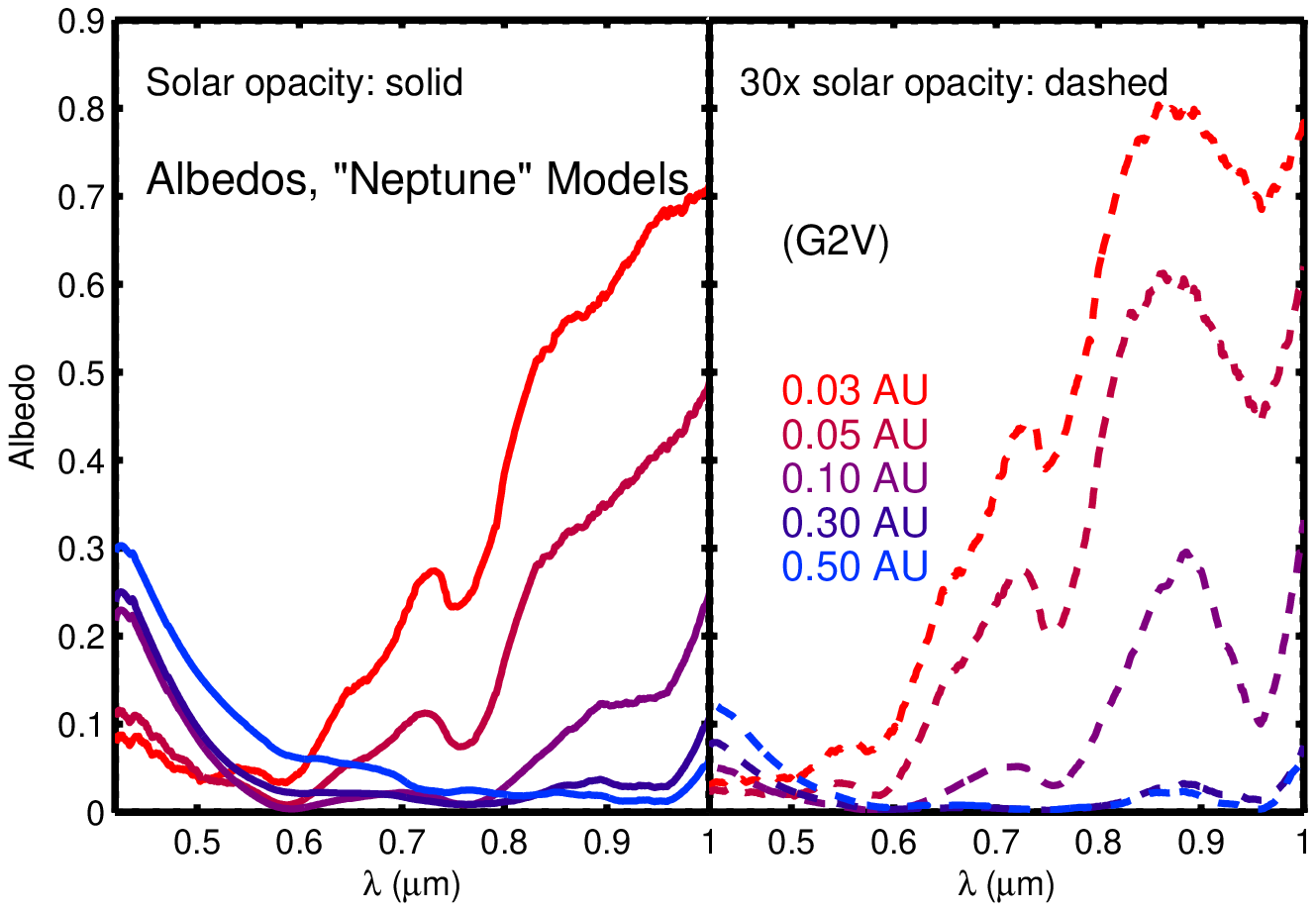}
}
\centerline{
\includegraphics[width=9.5cm,height=8cm,angle=0,clip=true]{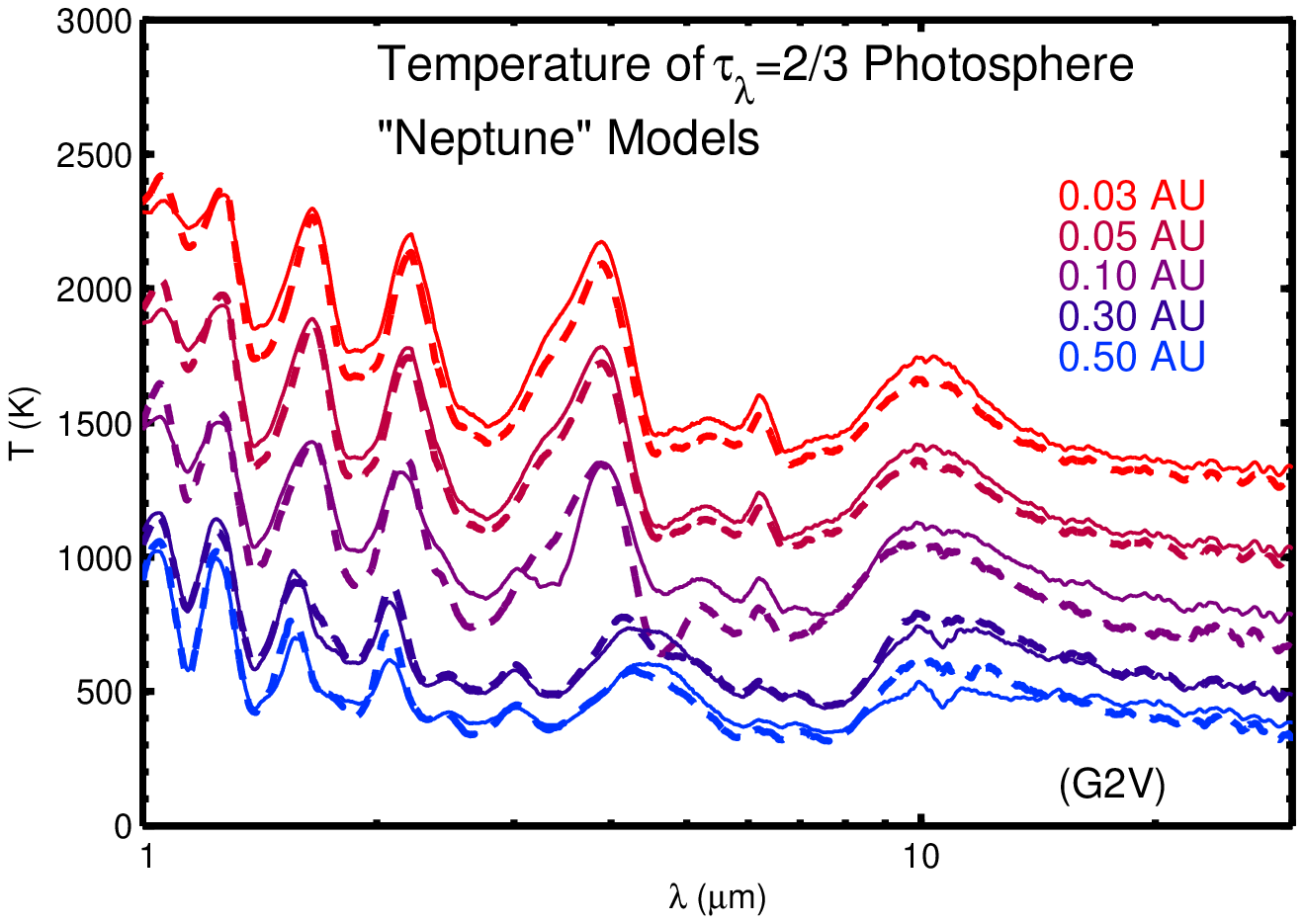}
\includegraphics[width=9.8cm,height=8cm,angle=0,clip=true]{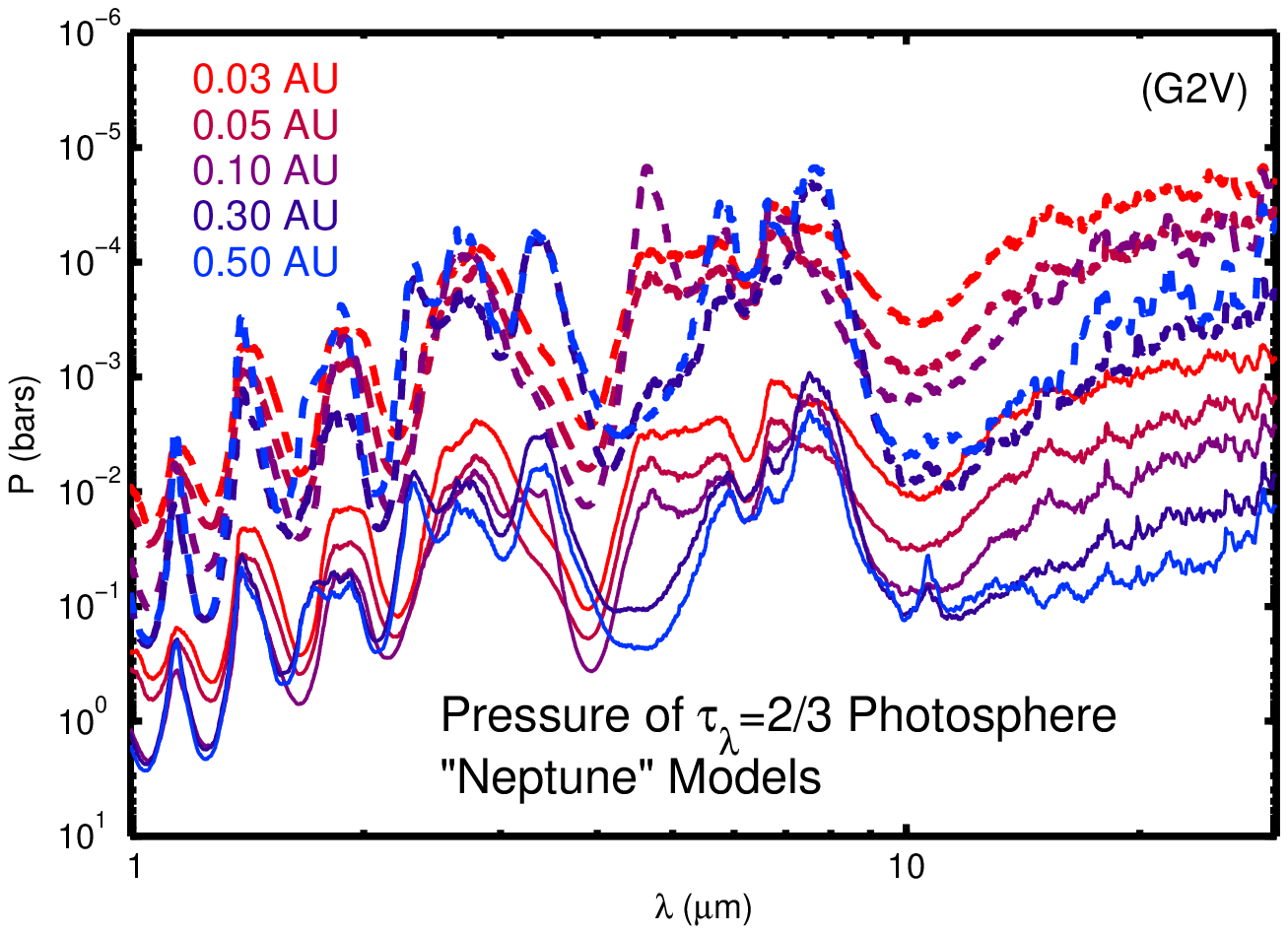}
}
\caption{Additional characteristics of models for different orbital
distances and metallicities depicted in the top panel of
Fig.~\ref{fig:dist_Z_stars_prof}. Analogous to
Fig.~\ref{fig:GJ436Pn_spec_alb_etc}. {\it Top left:} Planet-star flux
ratios for a GJ436b-like model planet at a variety of distances from
a G2V star, at solar and at 30-times solar metallicity.  Changes in
metallicity have only a very modest influence on model spectra.  {\it
Top right:} Optical albedos for these same models.  At high
metallicity, generally the cooler, more distant models (dashed curves)
have larger optical albedo.  At solar metallicity, the albedo varies
less smoothly with semimajor axis. {\it Bottom left:} Temperature of
the $\tau_\lambda=2/3$ surface, as a function of wavelength. {\it
Bottom right:} Pressure of the $\tau_\lambda=2/3$ surface, as a
function of wavelength.}
\label{fig:dist_Z_spec_alb}
\end{figure}

It was initially surprising to us that models with higher metallicity
around solar-type stars, with their concomitant changes in both
opacity and equation of state, nevertheless have extremely similar
spectral profiles to their lower metallicity cousins (solid and dashed
lines in top left panel of Fig.~\ref{fig:dist_Z_spec_alb}).  The
temperature-pressure profiles (top panel of
Fig.~\ref{fig:dist_Z_stars_prof}) of 30$\times$solar opacity models
are quite different from those of solar opacity models; increasing
metallicity appears to translate the $T$-$P$ profiles up to higher
altitudes (lower pressures).  But the photospheric temperature (bottom
panel of Fig.~\ref{fig:dist_Z_spec_alb}) remains essentially unchanged
with higher opacity.

How is this possible?  The photosphere pressure plot (lower right
panel of Fig.~\ref{fig:dist_Z_spec_alb}) offers the explanation.
Because, viewed from the outside, the optical depth increases so much
more quickly with depth into atmosphere for the high metallicity
models, their $\tau_\lambda=2/3$ surfaces are at significantly lower
pressures (by roughly an order of magnitude), corresponding to much
higher altitudes.  But, as Fig.~\ref{fig:dist_Z_stars_prof} shows,
higher in the atmosphere the temperature is higher in the high
metallicity models, and these two effects conspire to leave the
temperature of the photospheres of the high metallicity models
virtually unchanged (bottom left panel of
Fig.~\ref{fig:dist_Z_spec_alb}).

The changes to the albedos (top right panels of
Fig.~\ref{fig:dist_Z_spec_alb}) are not dramatic, although at the
short wavelength end the albedos of the 30$\times$solar metallicity
models are about half that of the solar metallicity models.
Furthermore, at the long wavelength end, a deep absorption feature due
to water arises at $\sim$0.95~$\mu$m in the high metallicity models
that is not present in the low metallicity ones.  Since the two
characteristics that are significantly different at high metallicity
(the $T$-$P$ profile and the photospheric pressure) are not
observable, these comparatively modest changes to albedo might
represent our best chance to constrain the atmospheric metallicity
through whole-disk observations (though transit spectrum observations
could provide a complementary constraint).  Note, however, that this
water feature is the same one seen in solar metallicity models of
GJ436b and HAT-P-11b in Figs.~\ref{fig:GJ436Pn_spec_alb_etc} and
\ref{fig:HP11Pn_spec_alb_etc}, and, therefore, might be a signature of
high metallicity only in the context of a modeling effort that
includes the appropriate stellar irradiance spectrum.

For a solar-type irradiation spectrum, the opacity (solar or
30$\times$ solar) of model planet's atmosphere has a minor influence
on the emergent spectrum.  However, the opacity has a stronger
influence when the irradiation spectrum is from a cool M-dwarf (as
shown in Fig.\ref{fig:GJ436Pn_spec_alb_etc}, where the irradiation is
from GJ436).

\subsection{Stellar Type}
\label{ssec:star}
How do a Neptune-mass planet's structure and appearance depend on its
host star?  Models T35a03Z01P0k0G through T62a03Z01P0k0G address this
question.  These models are of planets the size of GJ436b, separated
0.029~AU from stars ranging from cool (3500~K) to moderately hot
(6200~K).

\begin{figure}
\centerline{
\includegraphics[width=10cm,height=8.4cm,angle=0,clip=true]{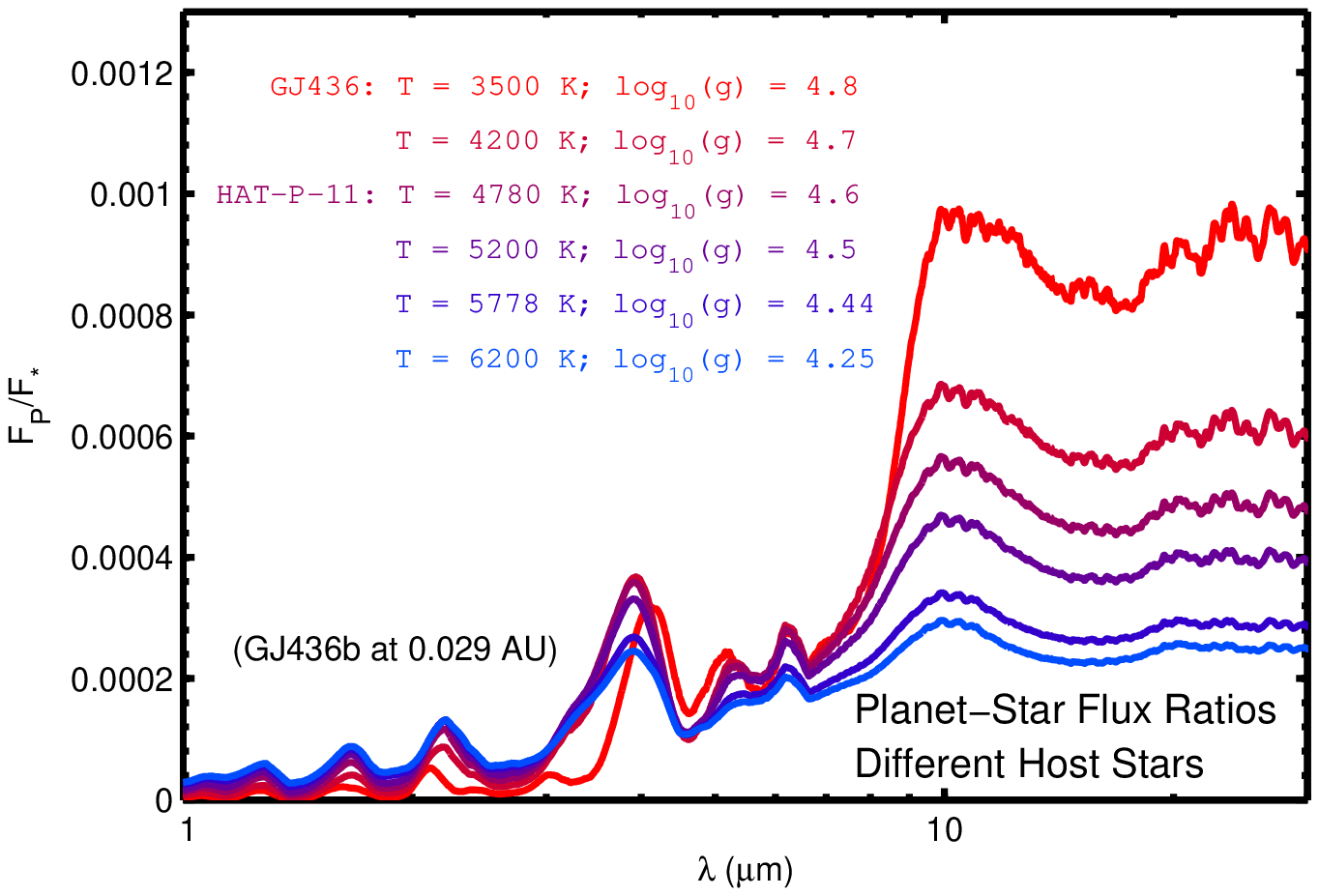}
\includegraphics[width=9.7cm,height=8cm,angle=0,clip=true]{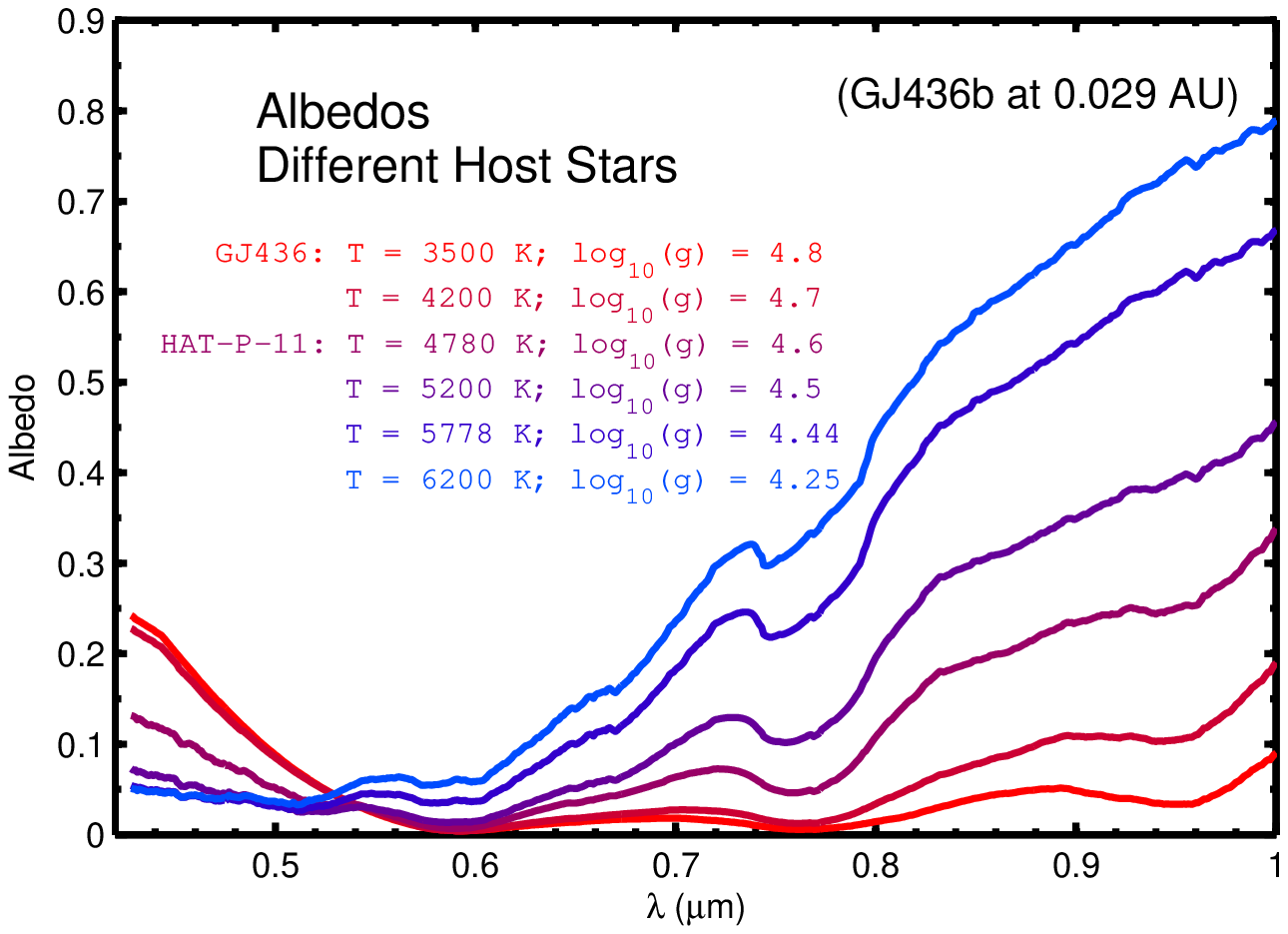}
}
\centerline{
\includegraphics[width=9.5cm,height=8cm,angle=0,clip=true]{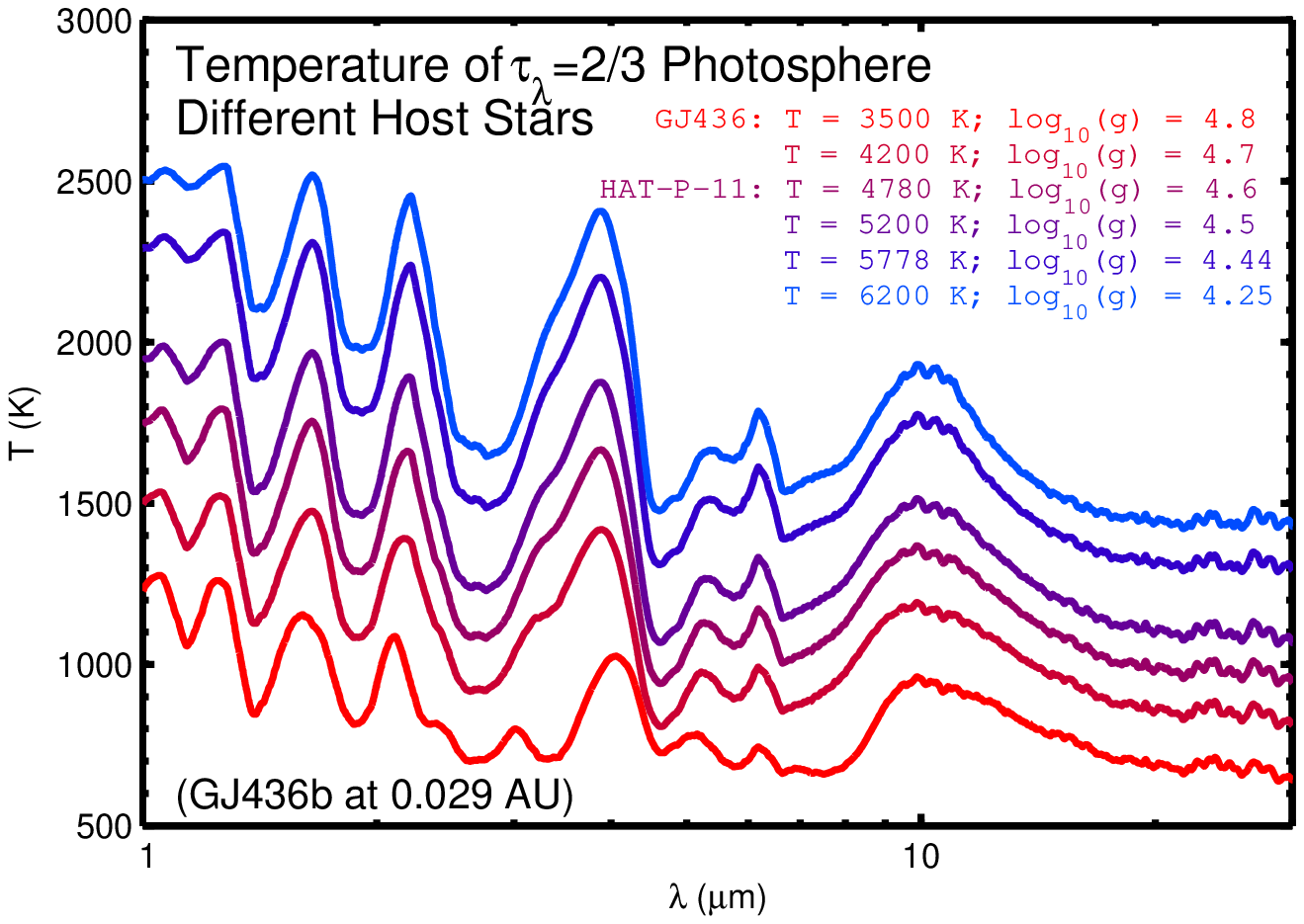}
\includegraphics[width=10cm,height=8cm,angle=0,clip=true]{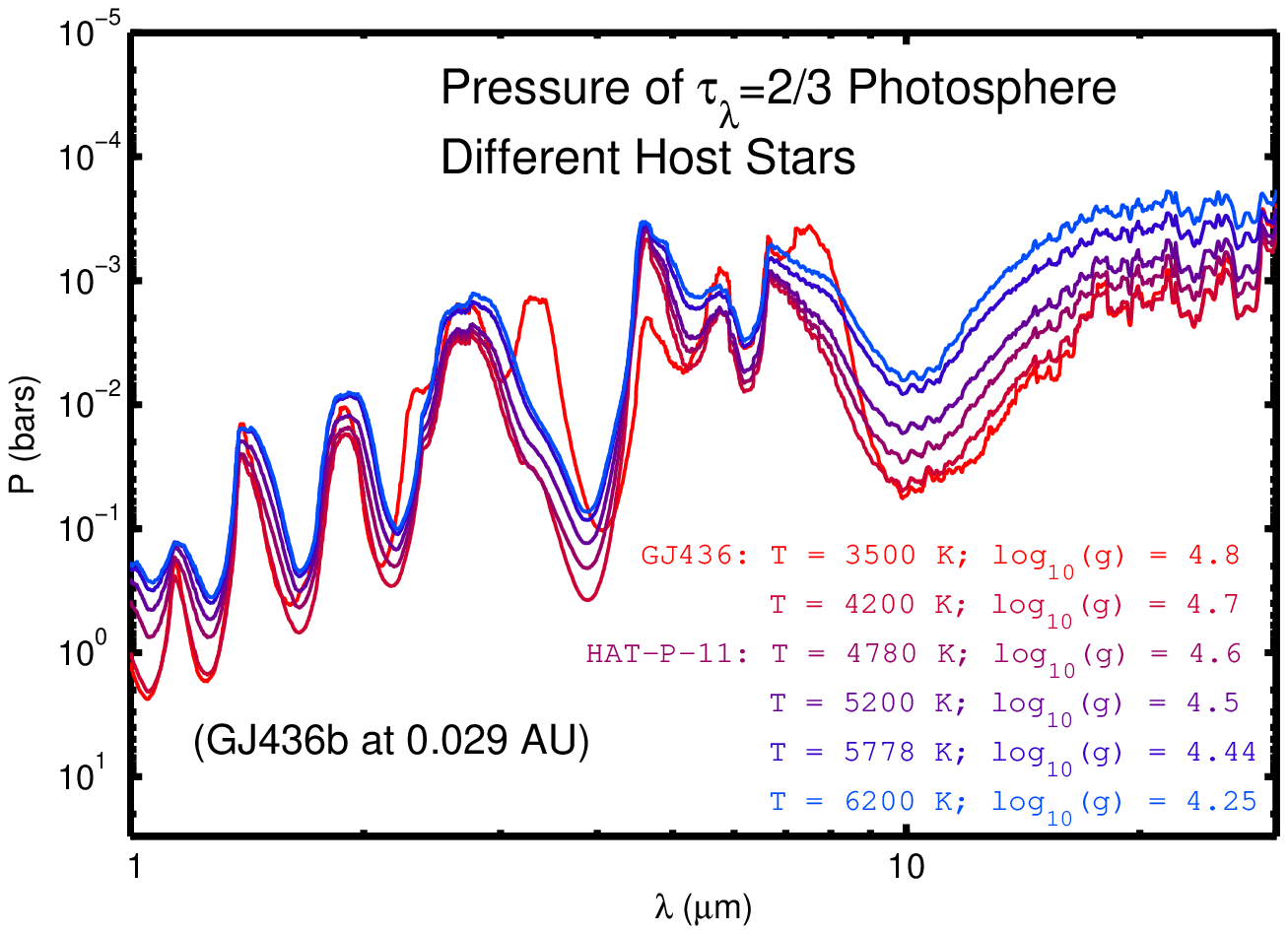}
}
\caption{Additional characteristics of models for different host stars
depicted in the bottom panel of
Fig.~\ref{fig:dist_Z_stars_prof}. Analogous to
Fig.~\ref{fig:GJ436Pn_spec_alb_etc}.}
\label{fig:Stars_spec_alb}
\end{figure}

The most prominent differences between models around cooler and dimmer
stars and those around hotter stars is that, unsurprisingly, planets
around hotter stars are hotter and bluer.  The bottom panel of
Fig.~\ref{fig:dist_Z_stars_prof} and the bottom left panel of
Fig.~\ref{fig:Stars_spec_alb} indicate how more massive, brighter
stars lead to hotter planets at a given distance.

For a planetary system that is of order a Gyr or older, a Neptune-mass
planet has lost almost all of its heat of formation
\citep{burrows_et_al2000, burrows_et_al2003, baraffe_et_al2005}.  As a
result, for planets at separations $\le$0.5~AU, contributions to the
emergent planetary flux (reflected/scattered and emitted) from
self-luminance are miniscule.  Therefore, the integrated planet flux
divided by the integrated stellar flux is, to a high degree of
approximation, simply $1/(4\pi)$ times the solid angle subtended by
the planet from the vantage of the star ($\Omega_p/4\pi = \pi{R_p}^2/
4\pi a^2$).  This leads to the somewhat counter-intuitive flux-ratios
plot in the upper left corner of Fig.~\ref{fig:Stars_spec_alb}: in the
mid-infrared (at wavelengths greater than $\sim$8~$\mu$m), a planet
around a cool star is significantly brighter relative to its star than
one around a hot star.  Of course, this is because the same total
relative flux is distributed predominantly at longer wavelengths for a
cooler planet and at shorter wavelengths for a hotter planet.

Interestingly, at wavelengths less than $\sim$8~$\mu$m, the
photospheric pressures are not dramatically different for the six
models, despite very different $T$-$P$ profiles.  Finally, similar to
the trend seen in albedo versus distance seen in \S\ref{ssec:dist},
cooler models have higher albedos at short wavelengths, while hotter
models have higher apparent albedos at longer wavelengths as the Wien
tail enters into the optical and near infrared.

\section{Summary and Conclusions}
\label{sec:conc}
We have presented a series of theoretical 1-D models of the radiative
and thermal structures of Neptune-mass, ``ice-giant'' exoplanets,
including models at opacities corresponding both to solar metallicity
and to 30$\times$solar.  To produce the latter, we computed
equilibrium chemical abundances of hundreds of molecular and atomic
species, and we recomputed the equation of state with the higher mean
molecular weight.  We produced models of the transiting planets GJ436b
and HAT-P-11b, and models of generic planets in a range of plausible
conditions.

In our investigation of GJ436b and HAT-P-11b, we find the following:
\begin{itemize}
\item The ratio of planet flux at 4.0~$\mu$m or at 10~$\mu$m to that
      at 4.6~$\mu$m might place a constraint on the amount of
      redistribution of heat from the day to the night side: more
      redistribution leads to a smaller flux ratio.  The ratio of
      planet flux at 7~$\mu$m to that at 4.6~$\mu$m might place a
      constraint on the possible presence of an extra optical absorber
      in the high atmosphere.
\item An extra optical absorber, if present in either planet, would
      have a more dramatic effect on both the $T$-$P$ profile and the
      emergent spectrum of HAT-P-11b than of GJ436b, because HAT-P-11
      is brighter in the optical.
\item Models of GJ436b without an extra absorber are significantly
      dimmer than the \citet{deming_et_al2007b} measurement, while the
      model with an extra absorber of opacity
      $\kappa~=~0.2\rm~cm^2~g^{-1}$ is closer to being consistent with
      the measurement, though still shy of it.
\item Models of these two planets have moderate optical albedos
      ($\sim$0.3) at the short wavelength end of the spectrum, and low
      albedos at longer wavelengths.
\end{itemize}

Furthermore, we find the following broad trends among generic models
of planets in this mass-range:
\begin{itemize}
\item Higher metallicity implies higher opacity for a given column
      mass of atmosphere.  As a result, photospheres of higher
      metallicity models are at lower pressures.
\item Nonetheless, when the irradiation spectrum is from a solar-type
      star, higher metallicity models appear to differ little from
      solar metallicity models in observable variables.  The most
      noticeable difference, among close-in planets around solar-type
      stars, is the water absorption feature at $\sim$0.95~$\mu$m that
      shows up in higher metallicity models.  When the irradiation
      spectrum is from a cool M-dwarf (GJ436), the metallicity of the
      planet's atmosphere has a larger influence on the emergent
      spectrum.
\item Neptune-mass planets around cooler stars have larger flux ratios
      in the mid-infrared than ones at the same distance around hotter
      stars.  (This conclusion is germane to Jupiter-mass planets, as
      well.)
\end{itemize}

Close-in Neptune-mass planets present an observational challenge
because, with radii a factor of $\sim 3$ smaller than Jupiter-mass
planets, they are roughly an order of magnitude dimmer.  However,
observations of GJ436b have already been made with {\it Spitzer}, and
additional ones are planned.  For GJ436b, 3.6 and 4.5~$\mu$m IRAC
observations will help to quantify both the amount of redistribution
of heat from the day to the night sides and the possible presence of
an extra absorber in the high atmosphere.  Future observations with
new facilities will provide constraints on models of other planets in
this mass range, and will sharpen our understanding of this class of
objects.

\acknowledgments

We thank Jason Nordhaus, Nikole Lewis, and Adam Showman for helpful
discussions.  We thank our anonymous referee for a number of helpful
comments that improved the manuscript.  This study was supported in
part by NASA grant NNX07AG80G.  We also acknowledge support through
JPL/Spitzer Agreements 1328092, 1348668, and 1312647.

\newpage

\bibliography{biblio.bib}

\clearpage

\begin{table}[ht]
\small
\begin{center}
\caption{Characteristics of $\sim$Neptune-Mass Exoplanets} \label{tab:neptunes_data}
\begin{tabular}{l|cccc|c|l}
\hline
\multirow{2}{*}
 {Planet}      & {Mass}                           & {Radius}                         & $\log_{10}g$ & {Semi-major Axis} & \multirow{2}{*} {Stellar Type} & \multirow{2}{*} {References} \\
               & ($M_{\rm Nep}$\tablenotemark{a}) & ($R_{\rm Nep}$\tablenotemark{b}) & (cgs)        & (AU)              &                                &                              \\

\hline
\hline
\rule {-3pt} {10pt}
CoRoT-7b       & 0.65            & 0.43            & 3.59   & 0.02              &  K0 V                          & \citealt{fressin_et_al2009}  \\[0.2cm]
GJ436b        & 1.33            & 1.04            & 3.15   & 0.03              &  M2.5                          & \citealt{torres2007}         \\[0.2cm]
\multirow{2}{*} {HAT-P-11b} & \multirow{2}{*} {1.50} & \multirow{2}{*} {3.05} & \multirow{2}{*} {1.22} & \multirow{2}{*} {0.05}  & \multirow{2}{*} {K4} & \citealt{bakos_et_al2009}   \\[0.2cm]
                            &                        &      &                        &                         &                      & \citealt{dittman_et_al2009} \\[0.2cm]
\hline
\rule {-3pt} {10pt}
HD 181433b     & 0.44            & -               & -      & 0.08              &  K3 IV                         & \citealt{bouchy_et_al2009}   \\[0.2cm]
HD 285968b     & 0.49            & -               & -      & 0.07              &  M2.5 V                        & \citealt{forveille_et_al2009} \\[0.2cm]
HD 40307d      & 0.53            & -               & -      & 0.13              &  K2.5 V                        & \citealt{mayor_et_al2009}    \\[0.2cm]
HD 7924b       & 0.54            & -               & -      & 0.06              &  K0 V                          & \citealt{howard_et_al2009}   \\[0.2cm]
HD 69830b      & 0.61            & -               & -      & 0.08              &  K0 V                          & \citealt{lovis_et_al2006}    \\[0.2cm]
HD 160691c     & 0.62            & -               & -      & 0.09              &  G3 IV-V                       & \citealt{pepe_et_al2007}     \\[0.2cm]
55 Cnc e       & 0.63            & -               & -      & 0.04              &  G8 V                          & \citealt{poveda_et_al2008}   \\[0.2cm]
GJ 674b        & 0.69            & -               & -      & 0.04              &  M2.5                          & \citealt{bonfils_et_al2007}  \\[0.2cm]
HD 69830c      & 0.70            & -               & -      & 0.19              &  K0 V                          & \citealt{lovis_et_al2006}    \\[0.2cm]
OGLE-05-169L b & 0.74            & -               & -      & $\sim$2.7         &  -                             & \citealt{gould_et_al2006}    \\[0.2cm]
HD 4308b       & 0.87            & -               & -      & 0.11              &  G5 V                          & \citealt{otoole_et_al2009b}  \\[0.2cm]
GJ 581b        & 0.91            & -               & -      & 0.04              &  M3                            & \citealt{mayor_et_al2009b}   \\[0.2cm]
HD 190360c     & 1.06            & -               & -      & 0.13              &  G6 IV                         & \citealt{vogt_et_al2005}     \\[0.2cm]
HD 69830d      & 1.08            & -               & -      & 0.63              &  K0 V                          & \citealt{lovis_et_al2006}    \\[0.2cm]
HD 219828b     & 1.22            & -               & -      & 0.05              &  G0 IV                         & \citealt{melo_et_al2007}     \\[0.2cm]
HD 16417b      & 1.28            & -               & -      & 0.14              &  G1 V                          & \citealt{otoole_et_al2009}   \\[0.2cm]
HD 47186b      & 1.33            & -               & -      & 0.05              &  G5 V                          & \citealt{bouchy_et_al2009}   \\[0.2cm]
HD 99492b      & 2.02            & -               & -      & 0.12              &  K2 V                          & \citealt{marcy_et_al2005}    \\[0.2cm]
HD 11964b      & 2.04            & -               & -      & 0.23              &  G5                            & \citealt{wright_et_al2009}   \\[0.2cm]
HD 49674b      & 2.13            & -               & -      & 0.06              &  G5 V                          & \citealt{wright_et_al2007}   \\[0.2cm]
55 Cnc f       & 2.67            & -               & -      & 0.78              &  G8 V                          & \citealt{fischer_et_al2008}  \\[0.2cm]
  \tableline
\end{tabular}
\bigskip
\tablenotetext{a}{The mass of Neptune is $M_{\rm Nep} = 1.02\times
10^{33} {\rm~ g} = 0.054~M_{\rm Jup}$, where $M_{\rm Jup} = 1.90\times
10^{33} \rm~g$.}
\tablenotetext{b}{The radius of Neptune is $R_{\rm Nep} = 2.48\times
10^9 {\rm~cm} = 0.346~R_{\rm Jup}$, where $R_{\rm Jup} = 7.15\times
10^9 \rm~cm$.}
\end{center}
\end{table}

\clearpage

\begin{table}
\small
\begin{center}
\caption{Exoplanet Models\tablenotemark{a}}
\label{tab:models}
\begin{tabular}{llcccccc}
\hline
\multirow{2}{*} {Model Name} & \multirow{2}{*} {Stellar Model}        & {$a$ } & {Metallicity} & \multirow{2}{*} {$P_n$} & $R_p $          & $\log_{10} g$ & $\kappa$          \\
                             &                                        & (AU)   & ($Z_\sun$)   &                         & ($R_{\rm Nep}$) & (cgs)    & (cm$^2$~g$^{-1}$) \\
\hline
\hline
\rule {-3pt} {10pt}
T35a03Z01P0k0G & GJ436   ($T=3500$ K; $\log_{10} g = 4.80$) & 0.03   & 1            & 0.0                     & 1.04            & 3.145    & 0.0     \\[0.2cm]
T35a03Z01P1k0G               & GJ436   ($T=3500$ K; $\log_{10} g = 4.80$) & 0.03   & 1            & 0.1                     & 1.04            & 3.145    & 0.0     \\[0.2cm]
T35a03Z01P2k0G               & GJ436   ($T=3500$ K; $\log_{10} g = 4.80$) & 0.03   & 1            & 0.2                     & 1.04            & 3.145    & 0.0     \\[0.2cm]
T35a03Z01P3k0G               & GJ436   ($T=3500$ K; $\log_{10} g = 4.80$) & 0.03   & 1            & 0.3                     & 1.04            & 3.145    & 0.0     \\[0.2cm]
T35a03Z01P4k0G               & GJ436   ($T=3500$ K; $\log_{10} g = 4.80$) & 0.03   & 1            & 0.4                     & 1.04            & 3.145    & 0.0     \\[0.2cm]
T35a03Z01P5k0G               & GJ436   ($T=3500$ K; $\log_{10} g = 4.80$) & 0.03   & 1            & 0.5                     & 1.04            & 3.145    & 0.0     \\[0.2cm]
\hline
\rule {-3pt} {10pt}
T48a05Z01P0k0H               & HAT-P-11 ($T=4780$ K; $\log_{10} g = 4.60$) & 0.05   & 1            & 0.0                     & 1.22            & 3.052    & 0.0     \\[0.2cm]
T48a05Z01P1k0H               & HAT-P-11 ($T=4780$ K; $\log_{10} g = 4.60$) & 0.05   & 1            & 0.1                     & 1.22            & 3.052    & 0.0     \\[0.2cm]
T48a05Z01P2k0H               & HAT-P-11 ($T=4780$ K; $\log_{10} g = 4.60$) & 0.05   & 1            & 0.2                     & 1.22            & 3.052    & 0.0     \\[0.2cm]
T48a05Z01P3k0H               & HAT-P-11 ($T=4780$ K; $\log_{10} g = 4.60$) & 0.05   & 1            & 0.3                     & 1.22            & 3.052    & 0.0     \\[0.2cm]
T48a05Z01P4k0H               & HAT-P-11 ($T=4780$ K; $\log_{10} g = 4.60$) & 0.05   & 1            & 0.4                     & 1.22            & 3.052    & 0.0     \\[0.2cm]
T48a05Z01P5k0H               & HAT-P-11 ($T=4780$ K; $\log_{10} g = 4.60$) & 0.05   & 1            & 0.5                     & 1.22            & 3.052    & 0.0     \\[0.2cm]
\hline
\rule {-3pt} {10pt}
T35a03Z01P3k2G               & GJ436   ($T=3500$ K; $\log_{10} g = 4.80$) & 0.03   & 1            & 0.3                     & 1.04            & 3.145    & 0.2     \\[0.2cm]
T48a05Z01P3k2H               & HAT-P-11 ($T=4780$ K; $\log_{10} g = 4.60$) & 0.05   & 1            & 0.3                     & 1.22            & 3.052    & 0.2     \\[0.2cm]
\hline
\rule {-3pt} {10pt}
T35a03Z30P0k0G               & GJ436   ($T=3500$ K; $\log_{10} g = 4.80$) & 0.03   & 30           & 0.0                     & 1.04            & 3.145    & 0.0     \\[0.2cm]
\hline
\rule {-3pt} {10pt}
T58a03Z01P0k0G               & G2 V     ($T=5778$ K; $\log_{10} g = 4.44$) & 0.03   & 1            & 0.0                     & 1.04            & 3.145    & 0.0     \\[0.2cm]
T58a05Z01P0k0G               & G2 V     ($T=5778$ K; $\log_{10} g = 4.44$) & 0.05   & 1            & 0.0                     & 1.04            & 3.145    & 0.0     \\[0.2cm]
T58a10Z01P0k0G               & G2 V     ($T=5778$ K; $\log_{10} g = 4.44$) & 0.10   & 1            & 0.0                     & 1.04            & 3.145    & 0.0     \\[0.2cm]
T58a25Z01P0k0G               & G2 V     ($T=5778$ K; $\log_{10} g = 4.44$) & 0.25   & 1            & 0.0                     & 1.04            & 3.145    & 0.0     \\[0.2cm]
T58a50Z01P0k0G               & G2 V     ($T=5778$ K; $\log_{10} g = 4.44$) & 0.50   & 1            & 0.0                     & 1.04            & 3.145    & 0.0     \\[0.2cm]
\hline
\rule {-3pt} {10pt}
T58a03Z30P0k0G               & G2 V     ($T=5778$ K; $\log_{10} g = 4.44$) & 0.03   & 30           & 0.0                     & 1.04            & 3.145    & 0.0     \\[0.2cm]
T58a05Z30P0k0G               & G2 V     ($T=5778$ K; $\log_{10} g = 4.44$) & 0.05   & 30           & 0.0                     & 1.04            & 3.145    & 0.0     \\[0.2cm]
T58a10Z30P0k0G               & G2 V     ($T=5778$ K; $\log_{10} g = 4.44$) & 0.10   & 30           & 0.0                     & 1.04            & 3.145    & 0.0     \\[0.2cm]
T58a25Z30P0k0G               & G2 V     ($T=5778$ K; $\log_{10} g = 4.44$) & 0.25   & 30           & 0.0                     & 1.04            & 3.145    & 0.0     \\[0.2cm]
T58a50Z30P0k0G               & G2 V     ($T=5778$ K; $\log_{10} g = 4.44$) & 0.50   & 30           & 0.0                     & 1.04            & 3.145    & 0.0     \\[0.2cm]
\hline
\rule {-3pt} {10pt}
T35a03Z01P0k0G\tablenotemark{b} & GJ436   ($T=3500$ K; $\log_{10} g = 4.80$) & 0.03   & 1            & 0.0                     & 1.04            & 3.145    & 0.0     \\[0.2cm]
T42a03Z01P0k0G               & M0 V     ($T=4200$ K; $\log_{10} g = 4.70$) & 0.03   & 1            & 0.0                     & 1.04            & 3.145    & 0.0     \\[0.2cm]
T48a03Z01P0k0G               & HAT-P-11 ($T=4780$ K; $\log_{10} g = 4.60$) & 0.03   & 1            & 0.0                     & 1.04            & 3.145    & 0.0     \\[0.2cm]
T52a03Z01P0k0G               & K0 V     ($T=5200$ K; $\log_{10} g = 4.50$) & 0.03   & 1            & 0.0                     & 1.04            & 3.145    & 0.0     \\[0.2cm]
T58a03Z01P0k0G\tablenotemark{c} & G2 V     ($T=5778$ K; $\log_{10} g = 4.44$) & 0.03   & 1            & 0.0                     & 1.04            & 3.145    & 0.0     \\[0.2cm]
T62a03Z01P0k0G               & F9 V     ($T=6200$ K; $\log_{10} g = 4.25$) & 0.03   & 1            & 0.0                     & 1.04            & 3.145    & 0.0     \\[0.2cm]
  \tableline
\end{tabular}
\bigskip
\tablenotetext{}{This table contains 31 rows describing the 29
models we computed.  Two rows are repeated.}
\tablenotetext{a}{The model name is a shorthand for the model
characteristics, as follows: ``T'' is for temperature, followed by the
temperature in Kelvin divided by 100; ``a'' is for semimajor axis,
followed by 10 times the orbital separation in AU; ``Z'' is for
metallicity, followed by the multiple of solar abundance; ``P'' is for
$P_n$, followed by $10\times P_n$; ``k'' is for $\kappa$, followed by
$10\times\kappa$ (in cm$^2$~g$^{-1}$).  The final letter is either
``G'' or ``H'': if ``G'', the planet's radius and surface gravity are
those of GJ436b; if ``H'', they are those of HAT-P-11b.}
\tablenotetext{b}{This is a repeat of the first row.}
\tablenotetext{c}{This is a repeat of the 15th row.}
\end{center}
\end{table}

\clearpage

\clearpage

\clearpage

\clearpage

\clearpage

\clearpage

\end{document}